\documentclass[twocolumn]{aastex631}
\usepackage{comment}

\begin{document}

\title{\texttt{Yuti}: A General-Purpose Transit Simulator for Arbitrary Shaped Objects Orbiting Stars}

\correspondingauthor{Ushasi Bhowmick}
\email{ushasibhowmick@gmail.com, ushasibhowmick@sac.isro.gov.in}

\author{Ushasi Bhowmick}
\affiliation{Space Applications Centre, Indian Space Research Organization, Ahmedabad, 380015, India}
\affiliation{Indian Institute of Space Science \& Technology, Thiruvananthapuram, Kerala 695547, India}

\author{Vikram Khaire}
\affiliation{Indian Institute of Space Science \& Technology, Thiruvananthapuram, Kerala 695547, India}
%%\affiliation{Physics Department, Broida Hall, University of California Santa Barbara, Santa Barbara, CA 93106-9530, USA}

\submitjournal{AJ}
\received{May 18, 2024}
\accepted{September 15, 2024}
% ---------------------------------------------------------------------------------
\begin{abstract}

We present a versatile transit simulator(\texttt{Yuti}) aimed at generating light curves for arbitrarily shaped objects transiting stars. Utilizing a Monte Carlo algorithm, it accurately models the stellar flux blocked by these objects, producing precise light curves. The simulator adeptly handles realistic background stars, integrating effects such as tidal distortions and limb darkening, alongside the rotational dynamics of transiting objects of arbitrary geometries. We showcase its wide-ranging utility through successful simulations of light curves for single and multi-planet systems, tidally distorted planets, eclipsing binaries and exocomets. Additionally, our simulator can simulate light curves for hypothetical alien megastructures of any conceivable shape, providing avenues to identify interesting candidates for follow-up studies. We demonstrate applications of \texttt{Yuti} in modeling a Dyson Swarm in construction, Dyson rings and Dyson disks, discussing how tidally locked Dyson disks can be distinguished from planetary light curves.
\end{abstract}

%----------------------------------------------------------------------------------------

%----------------------------------------------------------------------------------------
\section{Introduction} \label{sec:intro}

The discovery of the first exoplanet 51 Pegasi-b \citep{51peg}  in year 1995
has led to the emergence of numerous instruments on the ground and in space to identify and characterize exoplanets. With the launch of the {Kepler Space Telescope (Kepler)}  in 2009 \citep{kepmission, kepresult} followed by {Transiting Exoplanet Survey Satellite (TESS)} in 2018 \citep{tess}, the transit method became the most popular and efficient means to find exoplanets with more than 4000 exoplanet discoveries \citep{archive}. With the increasing sensitivity and precision of instruments, research has shifted from merely identifying exoplanets to studying detailed characteristics of them.

Transit photometry has uncovered signatures of many interesting phenomena beyond the detection of exoplanets and eclipsing binaries. This technique has been instrumental in identifying features such as starspots \citep[e.g.,][]{spotmodel_pop, kepstarspotstudy, wasp19spot}, and signatures of tidal interactions between host stars and exoplanets \citep[e.g.,][]{hatp13, wasp103b, hotjupiter_tides} leading to significant growth in the sub-field of Asteroseismology \citep[e.g.,][]{Seismology_surface_gravity, Seismology_kepredgiants, Seismology_TESS}. The study of transit timing variations has led to the discovery of further non-transiting planets \citep[e.g.,][]{ttvkepler, ttvkepler19, ttvhotjupiter}, orbital decay \citep[e.g.,][]{wasp12decay, wasp4decay} and apsidal precession - a gravitational phenomenon that causes the orbit of a planet to gradually rotate over time \citep[e.g.,][]{apsidecircimbinary, wasp19apside}.This effect, along with other findings like disintegrating planets \citep{brokenplanet, brokenplanet2}, exocomets \citep[e.g.,][]{exocomet_zero, exocomet_first, exocomet_recent} and exomoon candidates \citep[e.g.,][]{exomoon}, has expanded our understanding of the diversity and complexity of planetary systems.
Additionally, transit photometry has detected anomalous transit signals that have sparked interest in the search for technosignatures for the evidence of advanced civilizations \citep[e.g.,][]{tabbystar, tabbystar_laserline, transitanomaly, weirddetector}.

Technosignatures, which are evidence of technology that could not arise from natural processes and instead imply the presence of intelligent life, fall within the scope of the Search for Extraterrestrial Intelligence  \citep[SETI; e.g.,][]{Tarter01, Tarter07, brkthlisten}. These signatures encompass a wide range of indicators, from astroengineering projects and signals of planetary origin to interstellar space probes, offering ways to understand detectable features from extraterrestrial civilizations and potentially transforming our understanding of our place in the universe \citep[e.g.,][]{HMishra22, Wright22}.

%These signatures encompass a wide range of indicators, from astroengineering projects and signals of planetary origin to interstellar space probes, offering concrete evidence of extraterrestrial civilizations and potentially transforming our understanding of our place in the universe \citep[e.g.,][]{HMishra22, Wright22}.

Astroengineering projects, in particular, involve the construction of large-scale megastructures, with the Dyson Sphere \citep{fDyson_pap} being a prime example. This theoretical construct is envisioned as a massive shell surrounding a star to capture its energy output. The feasibility, sustainability, and potential designs of such megastructures have been explored in various studies \citep[e.g.,][]{dysph_dyswm, whitedwarf_dysph, wrightdy, dybreak}, alongside concepts like stellar engines \citep{steleng, steleng2} designed to propel stars. The presence of these megastructures could manifest as anomalies in the observed transit light curves, deviating from the patterns expected from spherical planets due to their unique geometries. Detecting such distortions requires careful analysis to distinguish them from natural phenomena affecting transit light curves. The role of transiting megastructures as key technosignatures in SETI research underscores the need for advanced techniques to identify unmistakable signs of extraterrestrial engineering \citep[see for e.g.,][]{megtranslc, spacemirrormention}.

Considering the wide range of observed transits and the speculative nature of megastructures, we have developed a versatile numerical tool designed to simulate the transit light curves of any objects orbiting stars. The simulator is called \texttt{Yuti}\footnote{https://github.com/ushasi-bhowmick/Yuti}, which is Sanskrit for conjunction, or transit. Our approach employs a Monte Carlo simulation to numerically generate transit light curves for transiting objects with arbitrary geometry. Previously, with a similar approach \cite{mctransit}, used Monte Carlo simulation to model transits and reflected light curves of non-spherical planets. \cite{eightbittransit} utilizes an 8x8 grid of pixels with varying opacity to identify different degeneracies in transit lightcurves. \cite{lcinversion} also uses a pixel-grid approach for shape reconstruction from lightcurve. Our simulator, in contrast, has no constraints on the resolution of the transiting geometry. In addition, our model can incorporate variations in different transit parameters.  \cite{exorings} and \cite{nonsphericaltransit} explore transit distortions due to exo-rings around planets. Such a geometry can be modeled using our simulator. Therefore our simulator aims to be a more general algorithm encompassing more specific geometries. This paper details our methodology for simulating transits, demonstrating its application across a variety of natural transiting phenomena and megastructure concepts. 

The paper is organized as follows. Section \ref{sec:TransitSim} 
describes the method used in \texttt{Yuti} and all its considerations. 
In section \ref{sec:NaturalTransit}, we study the performance of the simulator in modeling natural transits such as those of single planets, multi-planetary systems, systems experiencing tidal distortions, and exocomets. In section \ref{sec:AlienTransit}, we showcase the ability of our simulator to model a range of megastructures, including Dyson swarms, Dyson rings, and Dyson disks, with a particular focus on the detectability of Dyson disks. Finally Section \ref{sec:Summary} provides a summary of the paper's main results. 

%%########################################################################################

\section{Simulating transit light curve} \label{sec:TransitSim}

A transit occurs when an orbiting object comes in front of the star, with respect to the line of sight. The object blocks some light of the star; hence we observe a dip in the flux. Observations of multiple transits of the same object help us plot transit lightcurves in terms of its phase, where one transit period(i.e. orbital period) corresponds to a variation of phase angle($\theta$) from $-\pi$ to $\pi$.    
One widely used model for the study of transit lightcurves is provided in \citet{manda}. It models the transit of a planet assuming the planet as an opaque dark sphere and the star as a spherical body. This model describes the light curve of a transit using five parameters. These parameters include the relative size of the planet with respect to the star $R_{pl}/R_{st}$; the relative distance of the planet to the star i.e the semi-major axis of the orbit $R_{orb}$ with respect to the stellar radius ($R_{orb}/R_{st}$) and the impact parameter ($b$). The remaining two parameters $u_1$ and $u_2$ are limb darkening coefficients (see Eq.~\ref{eq:ldquad}) that are crucial for modeling the stellar limb darkening. 

Limb darkening is a phenomenon due to which a star appears brighter at the center and dimmer at the edges. There are many functional forms used to model the limb-darkening phenomena. However, in this work, we use the quadratic limb darkening law, 
\begin{equation}
\frac{I(\mu)}{I_0} = 1 - u_1 (1-\mu) - u_2(1-\mu)^2,
\label{eq:ldquad}
\end{equation}
and the non-linear limb darkening law \citep{claret_ld},
\begin{equation}
\frac{I(\mu)}{I_0} = 1 - a(1-\mu^{1/2}) - b(1-\mu) - c(1-\mu^{3/2}) - d(1-\mu^2).
\label{eq:ldnlin}
\end{equation}
Here, $I(\mu)$ is the intensity of the star at $\mu = cos \, \phi$ i.e, the cosine of the angular separation ($\phi$) of a point with respect to the center of the star, $I_0$ is the intensity at the center of the star, and $u_1, u_2, a, b, c$, and $ d$ are the limb darkening coefficients. We use the quadratic limb darkening law for the demonstration of our simulator, owing to its simplicity and the lesser number of free parameters. The non-linear limb-darkening law is used in Section 4.3.3, for analysing potential degeneracies in technosignature models. Other laws will be included in future versions of \texttt{Yuti}.

\subsection{Arbitrary Transit Model} \label{subsec:ArbTransit}

In order to extend the transit model to a generalized shape of the transit object, we need to invoke the basic principle of transit. The transit principle states that the relative change in the observed flux is proportional to the relative area of the blocking object with respect to the area of the star projected to the sky plane perpendicular to the line of sight, as given by
\begin{equation}
   \frac{\Delta F}{F} =  \frac{A_{pl}}{A_{st}}.
\end{equation}
Here $A_{pl}$ refers to the projected area of the blocking object overlapping with the background object, and $A_{st}$ refers to the projected area of the star or background object. An object with a bigger area of projection causes a deeper dip. In order to construct the light curve, we need to evaluate $\Delta F/F$ as a function of time ($t$) i.e, $A_{pl}(t)/A_{st}$. 
In the case of planets, for which spherical geometry is an excellent approximation, the projected area on the sky plane is always a circle. 
Therefore, it is possible to calculate  $A_{pl}(t)/A_{st}$ analytically for planets \citep[e.g. see][]{manda}. However, for an arbitrary object in transit, we have to switch to numerical methods to evaluate $A_{pl}(t)/A_{st}$. To do this, we utilize a standard Monte Carlo technique as described below. 

\begin{figure}
\plotone{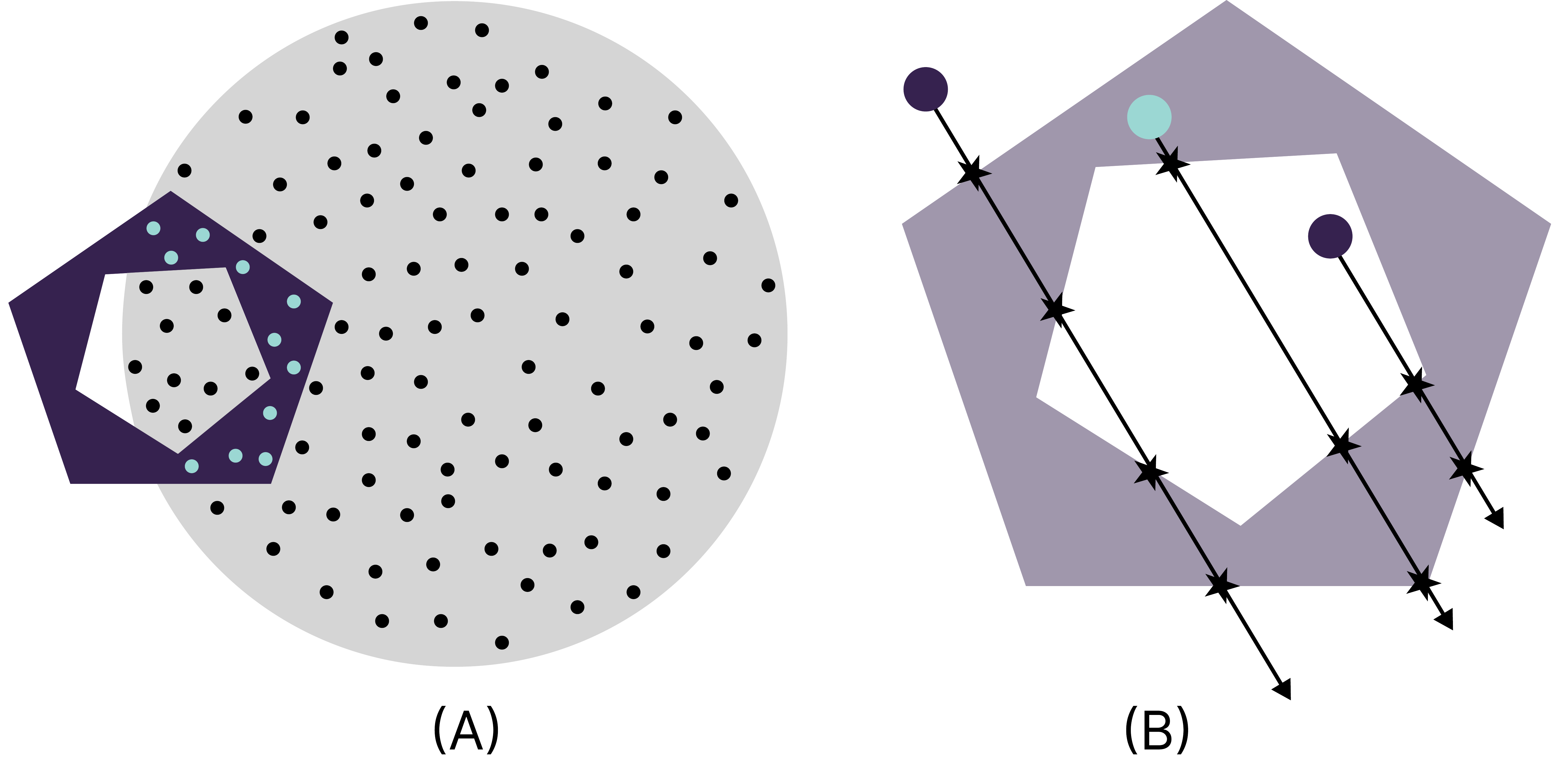}
\caption{Demonstration of the Monte-Carlo technique for numerical area calculation. (A) shows an outer circle and an inner shape, with each point having a certain probability of lying on the inside shape. A large number of sampled points is needed for an accurate estimate of the overlapped area between the background circle and foreground shape. (B) shows the ray casting technique, which is a method to evaluate whether a point lies inside or outside a particular shape by drawing a ray in any direction and calculating its intersection with the edges of the shape. An even number of intersections imply that the point is outside the shape, and an odd number of intersections means that the point is inside the shape.
\label{fig:mcpoints}}
\end{figure}

Consider an arbitrary background stationary shape (a circle), and a moving foreground shape (a hollow pentagon), as shown in Fig.~\ref{fig:mcpoints} (A). If we sample a random point on the circle, the probability of the point being placed within the transiting shape is equal to the ratio of the overlapping area of the shape ($A_{pl}$) to the background shape ($A_{st}$). This is given as, 

$$P = \frac{A_{pl}}{A_{st}}$$

\begin{figure}[ht]
\plotone{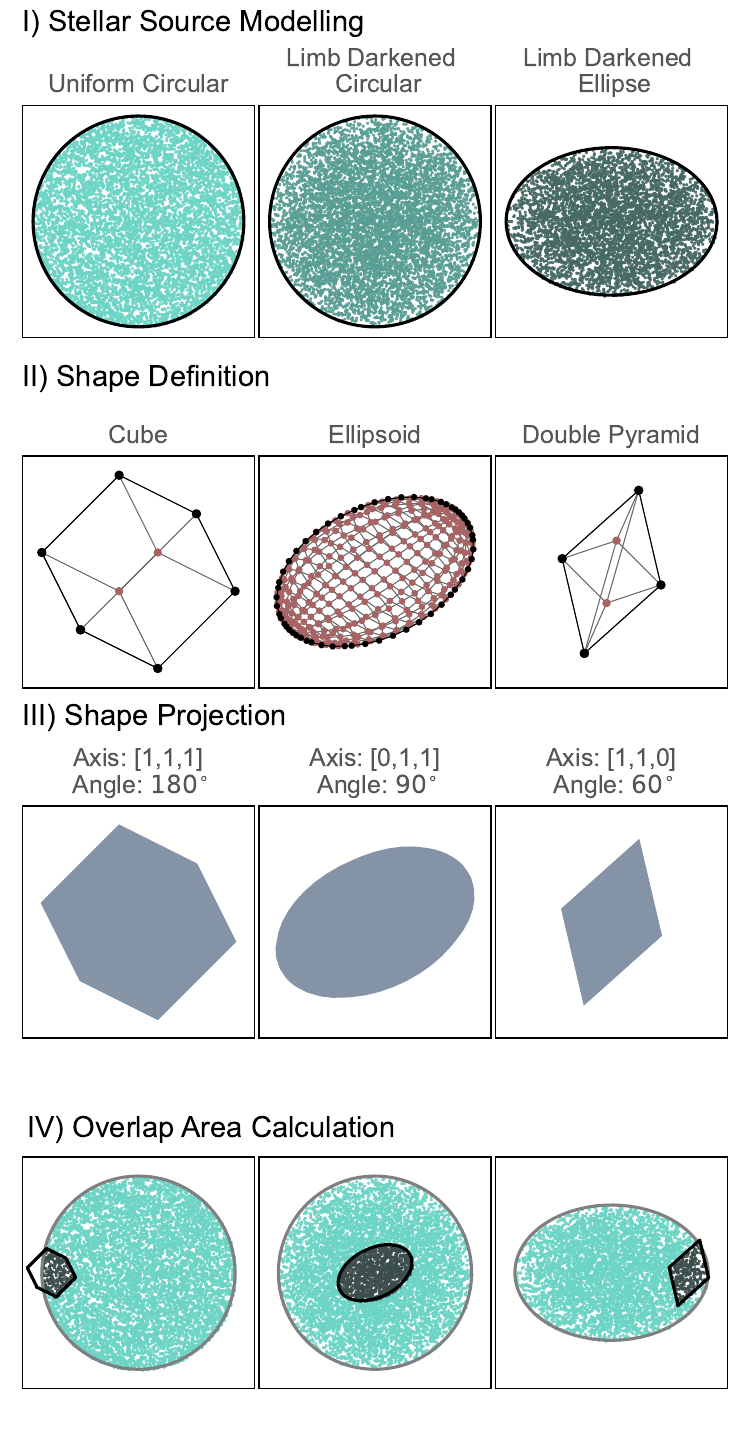}
\caption{Workflow of the transit simulation. (I) shows the modeling of the stellar source with a uniform, limb-darkened, and limb-darkened elliptical distribution.  (II) shows the definition of three arbitrary shapes (cube, ellipsoid, double pyramid) rotated by a random angle around an axis, using corners and edges. (III) shows the subsequent projected sub-shape. In (IV), the sub-shape is put in orbit around the star and the points inside the sub-shape are highlighted. The fraction of in points to out is used to calculate the area of overlap. 
\label{fig:mcdemo}}
\end{figure}

To evaluate $A_{pl}/A_{st}$, we need to evaluate $P$. To achieve this, we sample a number of random points inside the background shape. Then we count the number of points that simultaneously lie inside the inner shape. The fraction of such points gives us the value of $P$. This in turn gives us  $A_{pl}/A_{st}$ and therefore the normalized light curve $\Delta F/F$. The following subsections describe the exact algorithm, which is also demonstrated graphically in Fig. \ref{fig:mcdemo}.

%_______________________________________________________________
%______________________________________________________________
%_______________________________________________________
%_______________________________________________________

\subsection{Modelling a stellar source with limb darkening} \label{subsec:StellarSourceModel}

In this work, we assume the shape of the background star to be a sphere with the flexibility to include a more general case of a triaxial ellipsoid star. Even though the analogy of arbitrary shapes of transiting objects can be extended to arbitrarily shaped luminous objects as well, for the purpose of this work, only sphere and triaxial ellipsoids have been considered for modeling the stellar source.

We need to sample a large number of uniform random points within the projected area of the star on the sky plane i.e a circle. This requires two set of random numbers that represents polar coordinates, $r \in (0, R_{st}$) and $\theta \in (0, 2\pi)$. To generate a uniform distribution over a circular area, the distribution over $\theta$ is uniform. For $r$, we require a powerlaw distribution with index $0.5$, given as,

$$r = R_{st}\sqrt{x}$$

where $x$ represents a random number with uniform distribution. This is shown in the first panel in Fig. \ref{fig:mcdemo} (I).

To include limb darkening, the distribution must follow the intensity pattern prescribed by the chosen limb darkening law. This is done using the inversion method. For example, we select the quadratic limb darkening law (Eq.~\ref{eq:ldquad}) and convert the intensity to a cumulative distribution function (CDF) in terms of the radial distance $r$ of a random point on the circle of radius $1$,

\begin{eqnarray}
    CDF = \frac{1}{k} [ (1-u_1-2u_2)r^2 + \frac{1}{2} {u_2 r^4} + \nonumber \\ \frac{2}{3}(u_1+2u_2) \{ (1-r^2)^{3/2} -1\} ],
\end{eqnarray}
where,
\begin{equation}
    k = 1 - \frac{u_1}{3} - \frac{u_2}{6}.
\end{equation}
To obtain the correct distribution for $r$, we need to invert the CDF, by generating a uniform distribution in $(0,1)$ and solving for $r$. This in turn is scaled according to $R_{st}$. The resulting distribution of random points obtained using the above prescription is illustrated in the second panel of Fig. \ref{fig:mcdemo}(I). The sampling points show less density at the edges representative of the dimmer edges of the projected star to mimic the limb darkening. 

As an extension of the algorithm, we also describe a more general stellar shape: a triaxial ellipsoid with three axes $r_a, r_b, r_c$. We scale the circular limb darkened distribution with $r=1$ according to $\theta_s$ (between $r_1$ and $r_2$, the two axes of the projection). This gives us an elliptical projection as demonstrated in the right-hand panel of Fig. \ref{fig:mcdemo} (I).

\subsection{Modelling the Orbit} \label{subsec:OrbitModel}

To calculate the trajectory of the transiting object, we need to set the orbital parameters. This includes an orbital distance ($R_{orb}/R_{st})$ and an inclination towards the observer's line of sight. We sample the total phase $\theta = (0, 2\pi)$ with a chosen resolution (also called frame resolution in this work). This phase angle ($\theta$) is a measure of a fraction of the orbital period.  For circular orbits with constant angular velocity, the phase angle is equal to the orbital phase angle ($\Theta$). The orbital phase angle is a phase value corresponding to the location of the planet in its orbit at a certain time.

In order to model an eccentric orbit instead of a circular one, we use the Kepler equation for the distance (r) of the object  with respect to the star in the orbit, which is given by,

\begin{equation}
    r = \frac{a(1-e^2)}{1+e\, cos\Theta},
\end{equation}
where $e$ is the eccentricity of the orbit and $a$ is the semi-major axis length. For eccentric orbits, the angular velocity of the orbiting body is not constant, therefore the phase angle $\theta$ is not equal to the orbital phase angle $\Theta$. We account for this by calculating the orbital phase angle ($\Theta$) for each phase point using equations given 
below:
\begin{equation}
    \Omega = \frac{2\pi a^2 \sqrt{1-e^2}}{r^2},
\end{equation}

$$\Omega = \frac{\Delta \Theta}{\Delta t}, \quad \frac{\Delta \theta}{2\pi} = \frac{\Delta t}{P}.$$

Here $\Omega$ is the orbital velocity and $P$ refers to the orbital period of the system. Using $\Omega$ for each point, we can use the equations given above and calculate the $\Theta$ for each value of $\theta$ from $0$ to $2\pi$. 
Additionally, we also need the periapsis offset i.e. an angle that determines the orientation of the semi-major axis with respect to the line of sight.This along with $\Omega$ for each point allows us to determine the position of the object within its orbit.

\subsection{Modelling of Transiting Object} \label{subsec:TransitObjectModel}

We define a shape for the transiting object in terms of the coordinates of its corners. Arbitrary 3D shapes are defined by a set of three dimensional coordinates and their connected edges. We calculate the projection of the shape to our line of sight. We evaluate the projection of the object at each point in transit to generate a set of 2D sub-shapes at each point in the transit. If our shape also includes rotation, we alter coordinates accordingly by making use of rotation matrices. This is shown in Fig. \ref{fig:mcdemo} (rows II and III), where we take three different shapes, rotate them, and calculate their projections. The projected sub-shape is a list of those corners which form the outer boundary of the projection. These sub-shapes are then subjected to a Monte-Carlo sampling as described below.

\subsection{Monte-Carlo Simulation} \label{subsec:MCSim}

The sub-shape is translated to its position in the orbit. Then we calculate the overlapped area using the Monte-Carlo technique. For each of the randomly sampled points in the star, we evaluate whether it lies in the interior or the exterior of the shape. If the transiting object is a circle as in the case of planets, this amounts to evaluating the distance $d$ between the point in question from the center of the transiting circle. If the distance $d$ of the point is greater than the radius $r$ of the circle (i.e., $ d > r$) then the point is outside the circle, and if $d < r$ the point is inside the circle. For other shapes, we use a technique called ray casting to determine if the points are inside the shape or not.

Consider a random point and draw a ray extending from it in any (chosen) direction. If the point lies inside the shape, then the ray crosses the boundary of the curve an odd number of times (see panel B of Fig. \ref{fig:mcpoints}). If the point lies outside, it crosses the boundary for an even number of times. For our algorithm, we draw horizontal, right-seeking rays from each point and calculate their intersections to the edges of the shape. This determines whether each point lies in or out of the shape. The fraction of points lying inside the shape gives us the value of $\Delta F/F$ for that point in the trajectory, as is depicted in Fig. \ref{fig:mcdemo} (row IV). The methodology described above can be used to simulate the transit of any arbitrary shape orbiting a host star.

%%##########################################################################################

\begin{figure}%[ht]
\plotone{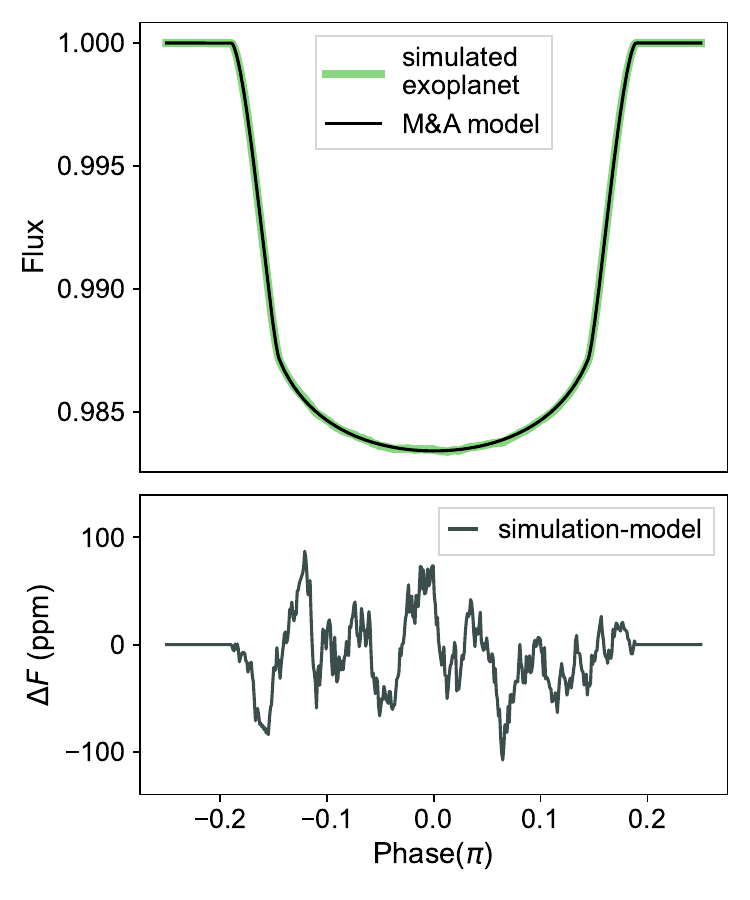}
\caption{Comparison of the simulation from \texttt{Yuti} with the standard model for a transiting exoplanet around a limb-darkened star. Simulated parameters are: $R_{pl}=0.12R_{st}$, $R_{orb}=2R_{st}$, $u_1=0.3$ and $u_2=0.2$. The top panel shows the simulation and the analytical model (overlapping) of \cite{manda}. The bottom panel plots the residuals (simulation - model) arising due to Monte-Carlo noise. The residuals are less than 0.5\% of the transit depth.
\label{fig:plsim}}
\end{figure}

\section{Modelling Natural Transits: Planets, Tidal Distortions, and Exocomets} \label{sec:NaturalTransit}
In this section, we validate our Monte Carlo technique used in \texttt{Yuti} by modeling several natural transit phenomena, including single and multi-planet transits, transits around tidally distorted stars, and exocomets. We compare our models to those reported in the literature, demonstrating that our technique is not only robust but also versatile. The following subsections present these examples.

\subsection{A Single Exoplanet Transit and Error Analysis} \label{subsec:simpleplanet}

To validate \texttt{Yuti}, we test it on a simple exoplanet transit. 
We simulate a planet with an arbitrary set of parameters (as $R_{pl}=0.12R_{st}$, $R_{orb}=2R_{st}$, $u_1=0.3$ and $u_2=0.2$) 
and compare it to the analytical transit model of \cite{manda} in Fig. \ref{fig:plsim}. Our simulation matches perfectly with the analytical model as shown in the top panel of Fig. \ref{fig:plsim} with overlapping curves. The difference between our simulation and the model is minimal, within $0.5\%$, as indicated in the residual plot in the bottom panel of Fig. \ref{fig:plsim}. The close match between the analytical model and our simulation confirms the accuracy and reliability of our technique. This minor difference arises from Monte-Carlo noise as explained below.

\begin{figure}[ht]
\plotone{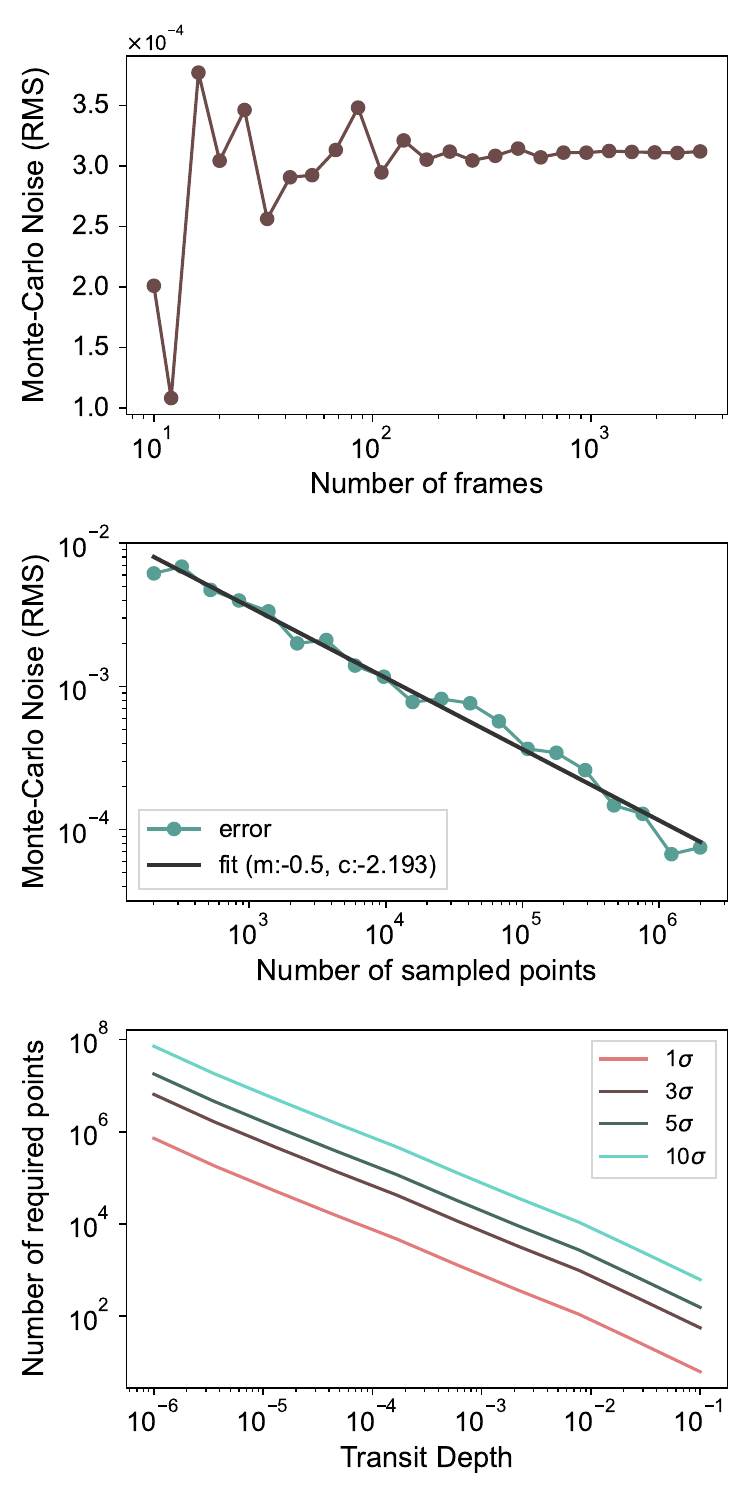}
\caption{Convergence of Monte-Carlo simulation. The top panel shows the variation of Monte-Carlo noise with changing frame resolution with a fixed sampling resolution of $10^5$. The RMS error flattens out to a fixed value of  $3.12 \times 10^{-4}$. The middle panel shows the variation of Monte-Carlo noise with sampling resolution, for a fixed frame resolution of $500$. The plot is linear in the log-log scale, as represented by the best-fit line ($y= mx + c$). The slope of this line is $m = -0.5$, which is characteristic of a Monte-Carlo distribution. The bottom panel shows the Monte-Carlo resolution required to simulate a system with a given transit depth at a required confidence level. Each of the solid lines represents a certain confidence level.
\label{fig:simperf}}
\end{figure}

The performance of the simulator depends on two factors, the first is the frame resolution i.e., the number of points used to model the orbit, and the second is Monte-Carlo resolution i.e., the number of points used to sample the star. 
The performance can be measured by evaluating the Monte-Carlo noise of the output light curve. To determine the Monte-Carlo noise, we calculate the root mean square (RMS) of the difference between two runs of the simulations performed with different sets of random numbers. We repeat this exercise for different frame resolutions and Monte Carlo resolutions. In Fig. \ref{fig:simperf}, we show the variation of Monte-Carlo noise with frame (top panel) and  Monte-Carlo resolutions (middle panel).
In the top panel of Fig. \ref{fig:simperf}, the Monte-Carlo resolution is fixed at $10^5$ points, while the frame resolution is variable. We observe that the RMS error flattens out to a fixed value of $3.12 \times 10^{-4}$ 
for a frame resolution greater than $500$ (i.e., $\Delta \theta < 0.0126)$ corresponding to the noise of the sampled distribution. In the middle panel, we keep the number of frames constant at $500$ and vary the Monte-Carlo resolution. The RMS error decreases with increasing number of sampled points following characteristic $1/\sqrt{n}$ decrement as illustrated by a fit to the RMS error in the middle panel of  Fig. \ref{fig:simperf} (the solid line).

The transit depth of a modeled exoplanet lightcurve varies greatly between hot Jupiters, which show deep transits, and super-earths which show very small transits. Therefore the precision required for simulating a particular system is strongly dependent on the size of the modeled system. The bottom panel of Fig. \ref{fig:simperf} shows the Monte-Carlo resolution required to model an object of a given transit depth at a required confidence. Each solid line represents a confidence level. For example, here, $3\sigma$ refers to a Monte-Carlo simulation where the ratio of the transit depth to the Monte-Carlo noise is $3$.  The vertical displacement of the lines show that more number of Monte-Carlo points are needed to sample the lightcurves at higher resolution. As we go to smaller planets with smaller transit depth, the number of points needed increases. However, the increase is logarithmic, and therefore implies a significant increase in computation requirement. 

In conclusion, our error analysis demonstrates that while the frame resolution significantly impacts the precision of the simulated light curve, achieving a frame resolution above $500$ points or $\Delta \theta < 0.0126$ effectively minimizes Monte Carlo noise, stabilizing the RMS error. Additionally, the observed trend in RMS error reduction with increasing Monte Carlo resolution underscores the inherent efficiency of Monte Carlo simulations in modeling the light curves. The algorithm is able to model even shallow transits at high confidence, by selecting an appropriate Monte-Carlo resolution.

%___________________________________________________________________________

%__________________________________________________________________________

\subsection{A multi-planetary system: Trappist-I} \label{subsec:multiplanet}
\texttt{Yuti} is also capable of generating light curves for multi-planetary systems. As of March 2024, out of approximately $5600$ stars known to host exoplanets, around 900 have more than one confirmed planet. The stars with the most confirmed planets include our Sun, Kepler-90 \citep{kep90}, and Trappist-I \citep{trappist_disc, trappist_res}. To demonstrate the versatility of our transit simulator, we simulate the transit light curves for the Trappist-I multi-planetary system as described below.

Trappist-I is known to have seven confirmed exoplanets. In order to simulate the light curve for those we run our simulation for one orbital period of the slowest planet, Trappist-Ih. The sizes of the planet and their orbital distance are taken from \cite{trappist_disc}. We obtain quadratic limb darkening coefficients for the star from the VizieR Catalog \citep{vizier}. For convenience, at the start of the simulation, all planets are assumed to be aligned, and behind the star.
The top panel shows a schematic of one frame of the simulation. The sizes of the planets have been scaled up and the distance to each planet has been scaled down in this figure for better visualization. The bottom panel shows the normalized output light curve of this system for one phase duration of the orbit of Trappist-Ih. The noise seen at the bottom of the transit dips is due to Monte-Carlo noise. This example highlights the simulator's ability to faithfully reproduce the light curves of the intricate multi-planetary systems.

%_________________________________________________________________________
\begin{figure*}[ht]
\plotone{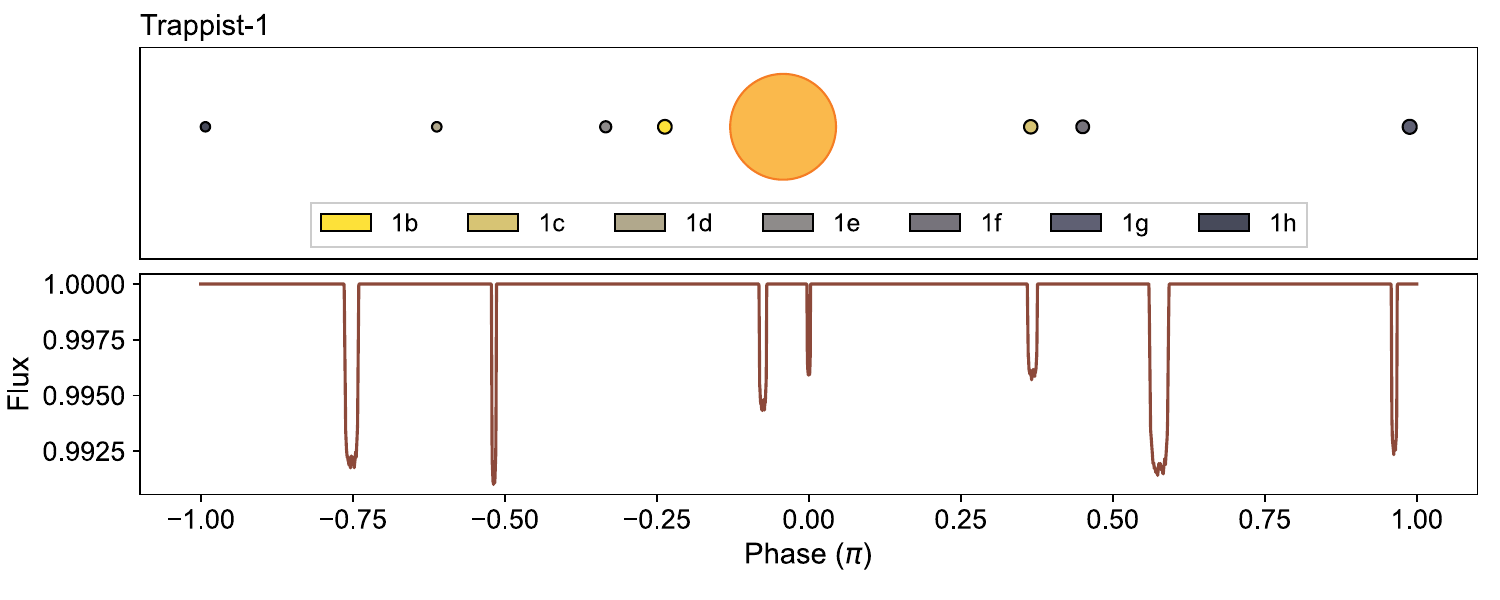}
\caption{Trappist-1 simulated light curve. The total phase duration is equal to one orbit of Trappist-1h. The top panel shows one frame of the simulation, with the seven planets. The size of the planets has been scaled up and distances have been scaled down for clarity. The planets have been color-coded by their orbital distance. The bottom panel shows the simulated light curve. The simulated light curve is for the realistic system.
\label{fig:trappist}}
\end{figure*}

%_________________________________________________________________________

%################################################################################

\subsection{Tidally Distorted Systems} \label{subsec:tides}
Tidal effects can cause significant distortions to the shape of the transiting object as well as the host star. Therefore, such systems lead to a variety of non-spherical as well as dynamic shapes which provide excellent grounds for testing \texttt{Yuti} against natural, observable, and already modeled phenomena. In this section, we simulate examples of tidally distorted systems and compare them to existing models.

\subsubsection{Tidally Distorted Exoplanet} \label{subsubsec:wasp103b}

Giant planets very close to their host stars, such as hot Jupiters,  are expected to undergo significant tidal distortions due to the gravitational effects of the star. Most of these systems are tidally locked and hence show a synchronization of their rotation and orbital period resulting in circularization of their orbit. 

The model for tidal distortion developed by \cite{correia} has been utilized to explain peculiar light curve features in many eclipsing binaries. Due to the large size, brightness, and mass, the tidal effects of the companion star on the host star are significant, which adds to the observed unique light curves. The same model can be extended to exoplanets, but in these cases, the tidal distortion signal is weak because of the small size, little brightness, and low mass of the planet compared to the host star. Moreover, a hot Jupiter is expected to have a negligible effect on the shape of the host star, which makes it more challenging to detect. 

Tidal distortion in hot Jupiters has been extensively studied \citep[e.g., ][]{wasp121, hatp13}. For example, WASP-4 \citep{wasp4decay} and WASP-12 \citep{wasp12decay} have been found to show a tidal decay. The first tidally distorted system to be observed at a significance level of $3\sigma$ is WASP-103b (\cite{wasp103b}). Due to the severity of the tidal effects of this system, 
it has been observed as a distortion in the shape of the transit. This system has been studied extensively by the Characterizing Exoplanet Satellite \citep[CHEOPS;][]{cheopsmission}.

Here, we demonstrate how \texttt{Yuti} can be used to model tidally distorted planets. For that, we choose the example of WASP-103b as a case study. The planet is modeled as a triaxial ellipsoid \citep[see e.g.,][]{correia} with axes $r_a, r_b, r_c,$ where:
\begin{equation}
    r_v = \sqrt[3]{r_a r_b r_c},
    \label{eq:tide1}
\end{equation}
with,
\begin{eqnarray}
    r_a=r_b(1+3q),     \nonumber \\           
    r_c=r_b(1-q),
    \label{eq:tide2}
\end{eqnarray}

and, the asymmetry parameter $q$ is given by
\begin{equation}
    q = \frac{h_f}{2} \frac{m_p}{m_s} \left(\frac{r_b}{r_0}\right)^3.
    \label{eq:tideq}
\end{equation}
Here, $r_v$ is the mean radius of the triaxial ellipsoid, $m_p/m_s$ represents the mass ratio of the planet to the star, $r_0$ is the distance between the planet and the star and $h_f$ is the second fluid love number. The second fluid love number is a parameter that describes the rigidity of a body under the effect of gravity. A completely fluid body has an $h_f$ value of $1$. Therefore, three parameters are required for fitting: $m_p/m_s$, $r_v$, and $h_f$. 

For the demonstration, we choose nine transits of WASP-103b from CHEOPS public data products (program CH\_PR100013). To obtain the orbital parameters by fitting this data, we utilize the model \emph{ellc} (\cite{ellc}) which utilizes the model by \cite{correia} to generate a light curve of the system. We use the fitted parameters given in \cite{wasp103b} to generate a model light curve for WASP-103b and compare it to our simulation. The values given in \cite{wasp103b} are $m_p/m_* = 0.001165$, $r_v = 0.1395$  and  $h_f = 1.59$.

For our simulation, we use the equations \ref{eq:tide1}, \ref{eq:tide2}, \ref{eq:tideq} and obtain the values for the three axes of the triaxial ellipsoid for WASP-103b as $r_a= 0.1345$, $r_b= 0.1258$, $r_c= 0.1605$. We define an ellipsoid planet with these radii and a spherical star with quadratic limb-darkening parameters $u_1 =0.5269$ and $u_2 = 0.1279$.

%_____________________________________________________________________
\begin{figure}[ht]
\plotone{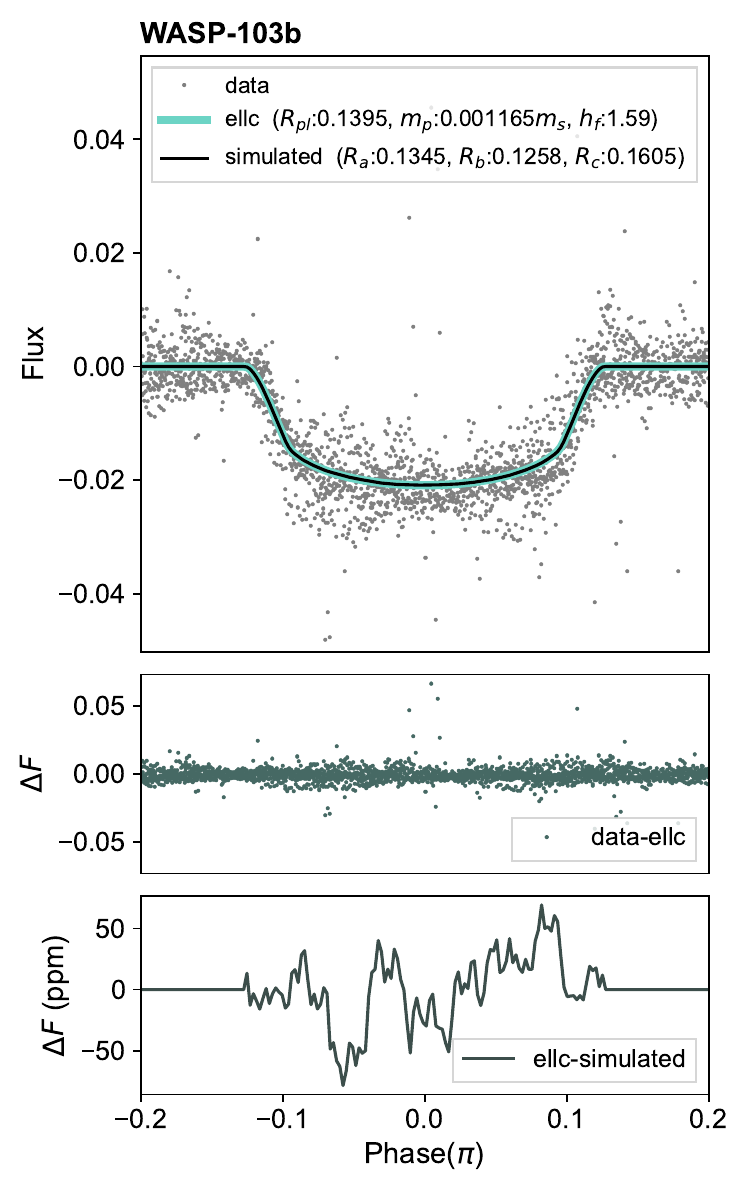}
\caption{Simulated light curves for WASP-103b. The top panel shows phase folded transits from CHEOPS (gray), the \emph{ellc} light curve with appropriate parameters, and the simulated light curve from \texttt{Yuti}. The middle panel shows the residuals from the observation (data-\emph{ellc}), whereas the last panel shows the residuals between the simulation and \emph{ellc}.
\label{fig:tidewasp}}
\end{figure}
%_____________________________________________________________________

%_____________________________________________________________________
\begin{figure*}[ht]
\plotone{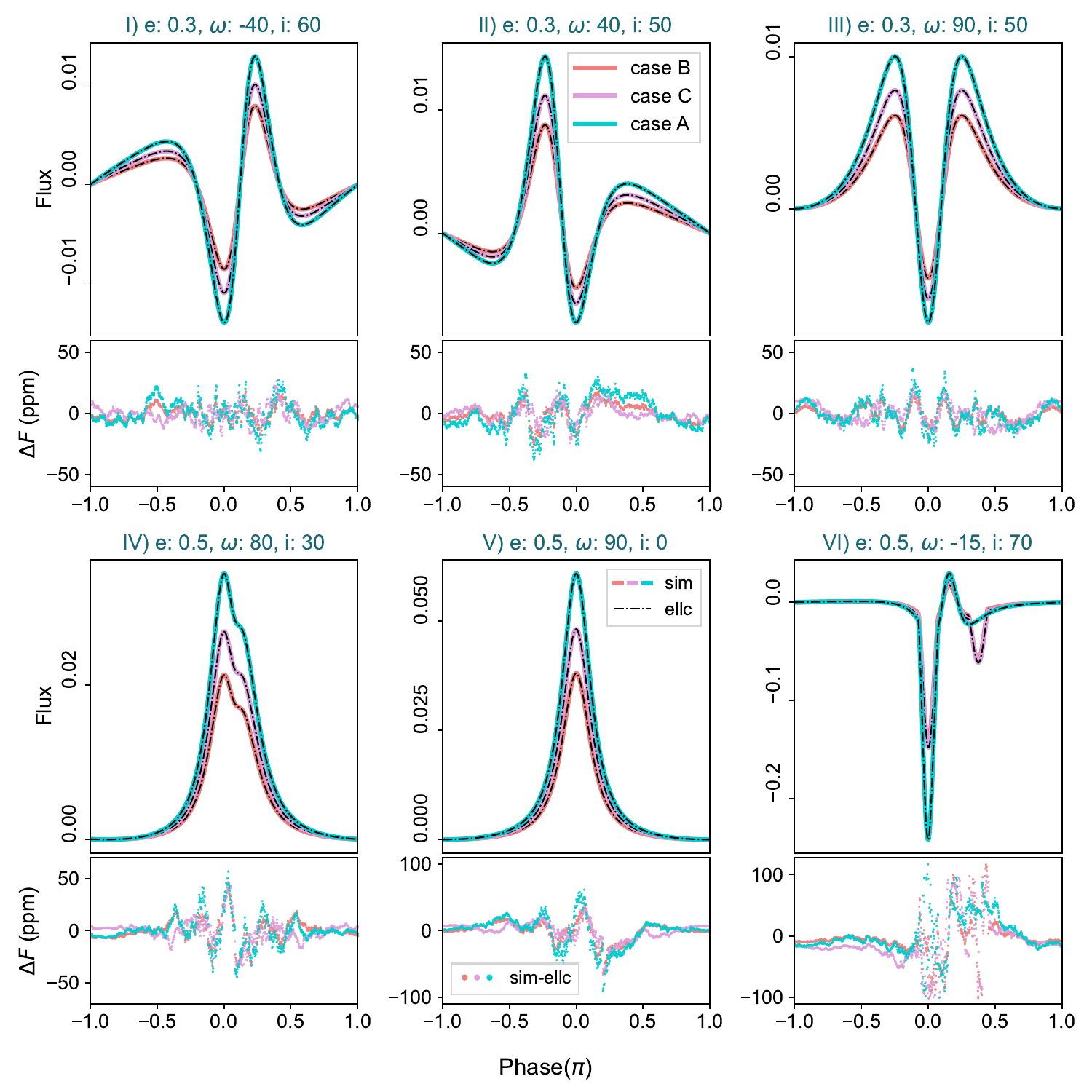}
\caption{Simulated heartbeat star systems in various configurations. Each panel is a configuration with a value of eccentricity, periapsis offset and inclination ($e, w, i$). The radius-ratio ($r_s/r_p=0.8$), semi-major axis ($5 r_p$), fluid love number ($1.5$) and mass ratio ($m_s/m_p=0.9$) is the same for all the configurations.  Three cases are considered. Case A represents an ellipsoidal star with spherical opaque companion (relative flux $f=0$, \emph{cyan curve}). Case B represents an ellipsoidal star with a spherical bright companion ($f=1$ \emph{red curve}). Case C considers an ellipsoidal star with an ellipsoidal bright companion ($h_f=1, f=1$, \emph{pink curve}). The top panels also show the \emph{ellc} model (\emph{black dot-dash line}) which coincides with the simulated lightcurves from \texttt{Yuti}. The bottom panels plot the residuals. Panels I-V represent heartbeat signatures at lower inclination that show no transit. Panel VI represents the highest inclination case which shows primary and secondary transits along with tidal deformation.
\label{fig:hbtide}}
\end{figure*}
%_____________________________________________________________________
 
The result of our simulation (black lines) is shown in Fig. \ref{fig:tidewasp} along with the observations from CHEOPS (gray data points).  We find that our simulated light curve agrees closely with the fit provided by \emph{ellc}. The middle panel shows the difference between the observed data and \emph{ellc} fit. In the bottom panel, we show the residuals between our model and the \emph{ellc} fits. The residuals in the bottom panel are far less than the residuals of the fit in the middle panel, which shows that our model is in excellent agreement to the data. The bottom panel residuals are of the order of 50 ppm, which is mainly because of the Monte Carlo noise.

This exercise demonstrates that \texttt{Yuti} is capable of reliably and accurately modeling the light curves of tidally distorted planets. In the subsequent subsection, we apply our simulator to model tidal distortions in highly eccentric eclipsing binaries known as heartbeat stars. 

\subsubsection{Heartbeat Stars} \label{subsubsec:heartbeattides}
Heartbeat stars, a class of eclipsing binary stars orbiting in highly eccentric paths, represent unique astrophysical systems where the eccentricity of their orbits induces significant distortions in their light curves. When the stars are at their closest approach, the proximity leads to tidal distortions and pulsations, causing the characteristic heartbeat shape of their light curves. A large number of heartbeat star lightcurves have been identified by \emph{Kepler} \citep[e.g.,][]{keplerhb1, keplerhb2} and \emph{TESS} \citep[e.g.,][]{tesshb}.

The shape of the light curve of the heartbeat system can vary widely. It depends primarily on three parameters, the eccentricity of the orbit ($e$), the longitude of periastron ($\omega$), and the inclination of the orbit ($i$). The top panels of Fig. \ref{fig:hbtide} show some of these configurations. 

To simulate heartbeat stars, we make use of the tidal equations in eq. \ref{eq:tideq}, where the asymmetry parameter $q$ depends on the distance to the center of the star.  We evaluate this distance for each point in the trajectory, which gives us values of $r_a$,  $r_b$, and $r_c$ (see Eq. \ref{eq:tide1}, \ref{eq:tide2}) for the ellipsoidal star (for illustration see the last panel in Fig. \ref{fig:mcdemo} I). 

For comparison, we take three cases of heartbeat systems. Case A considers an ellipsoidal primary star with a spherical secondary opaque companion (relative brightness = 0). Case B considers an ellipsoidal star with a bright secondary spherical companion (relative brightness = 1). Case C considers an ellipsoidal star with a bright ellipsoidal companion (relative brightness = 1). 
We choose an $h_f$ value of $1.5$ for the primary star. The size ratio of secondary to primary is taken as $0.8 r_s/r_p$, where $r_s$ refers to the radius of secondary companion, and $r_p$ refers to the radius of the primary star. The semi-major axis is taken as $5 r_p$. The mass ratio of secondary to primary is $0.9 m_s/m_p$, where $m_s$ refers to the mass of the secondary companion and $m_p$ refers to the mass of the primary star. Both coefficients of the quadratic limb darkening law are taken as $0$. For Case C, we choose $h_f = 1$ for the companion star. The choice of the parameters is made such that the difference in lightcurves between the three cases is maximized for better understanding. Similar to the model \emph{ellc}, for simplicity, we assume a synchronous rotation of the primary star, tidally locked to the secondary star. In reality, these systems do not exhibit perfect synchronous rotation \citep{hb_async_old, hb_async}.

The top panels of Fig.~\ref{fig:hbtide} show the simulated light curves of a heartbeat system in six different configurations. Panels I-V represent systems at lower inclination, which show tidal deformation signatures but no transit. Panel VI represents a configuration with a high inclination, and is the only one to exhibit transits. For comparison, we also generate a heartbeat light curves using the \emph{ellc} model with the same parameters. In the bottom panels, we show the difference between our model and model \emph{ellc} which are of the order of ppm and are a result of the Monte-Carlo noise. Some of the residuals in the bottom panels show fluctuations (see Fig. \ref{fig:hbtide} VI). These are due to narrow features with sharp transitions in the light curve caused by the high eccentricity and the resulting increase in orbital velocity near the heartbeat. This indicates that we need more frame resolution to reduce the noise. A refinement of the frame resolution near the sharp transitions would also help to capture these features more accurately.

We see that for the first five panels, Case A with opaque spherical companion shows the maximum amplitude in peaks and dips in the heartbeat. Case B with bright spherical companion shows the minimum amplitude in peak and dips. Due to the illumination of both objects, the relative change of flux due to distortion of the primary is diminished. In Case C, the distortion in the secondary as well as primary causes an increase in the total flux observed during the distortion, Therefore, the amplitude of peaks and dips are greater than case B. However, this is still less than case A, because of the illumination of the secondary companion in case C.

The effects are more complex in panel VI due to the occurrence of transits in addition to the heartbeat. Lightcurves for case B and C are nearly indistinguishable, however, case C is observed to show a higher amplitude than case B at the central peak. We see the transit of the primary star by the secondary at phase $0$ and the transit of the secondary star by the primary at phase $0.5$. In case A, this secondary transit at phase $0.5$ is not observed, because the eclipse of the opaque companion does not cause a dip. Also, the primary transit is substantially deeper, due to the opaque nature of the companion.

This analysis takes a simplistic representation of heartbeat stars. Ellipsoidal stars are expected to show gravity darkening due to the fast-rotating nature of the system \citep{gde_rot, gde_binary}. Also, eccentric systems show asynchronous or pseudo-synchronous rotation. We aim to include such effects in future versions of the simulator. Nevertheless, this exercise demonstrates that our simulator can be used to accurately model complex light curves of tidally distorted stars such as heartbeat systems. It emphasizes that our simulator can adeptly handle various intricate scenarios. Moving forward, in the next subsection we will simulate the light curves of a known exocomet candidate.

%################################################################################

\subsection{Exocomets} \label{subsec:exocomet}
The discovery of exocomets around $\beta$-Pictoris \citep{exocomet_zero} predates the discovery of exoplanets. With the launch of \emph{TESS} this comet system has been studied in extensive detail. Extensive modeling of an exocomet transit has been done by \cite{exocomet_first} and \cite{exocometbrogi}. These models along with \emph{TESS} data have been used to study the $\beta$-Pictoris system in great detail. This has also prompted the search for exocomet transits in existing Kepler light curves.

In this work, we adapt the model of \cite{exocomet_rappaport} which has been used to find exocomet candidates in the star KIC 3542116 and KIC 11084707. In this work, the exocomet in the star KIC 3542116 has been modeled as an occulter with sustained dust outflows. Because comets are expected to be in highly eccentric orbits, the orbital period of the comet is assumed to be much larger than the transit duration. Hence, the transverse velocity $v_t$ is assumed to be constant during the transit. The dust tail is also assumed to be narrow, and its extinction follows an exponential decay profile with the distance from the comet. The profile of the dust tail is given as:
\begin{equation}
    \tau = Ce^{-(\frac{x-x_c}{\lambda})}  \quad x<x_c; \, y-y_c<\Delta b/2
\end{equation}
Here, $\tau$ is the optical depth of the comet dust tail, $C$ is a normalization constant, $\lambda$ is the decay parameter for the optical depth, $x$ is the $x$-coordinate of a point on the stellar projection on the sky plane and $x_c$ is the $x$-coordinate of the center of the comet nucleus. The width of the comet and the dust tail perpendicular to its velocity is given as $\Delta b$ which is assumed to be negligible as compared to $R_{st}$ and therefore is degenerate with the normalization constant $C$. Therefore, the comet is modeled as six free parameters, $t_0, \lambda, C, v_t, b, DC$, where $b$ is the impact parameter of the transit, and $DC$ is the background flux level away from the transit.

\begin{figure}%[ht]
\plotone{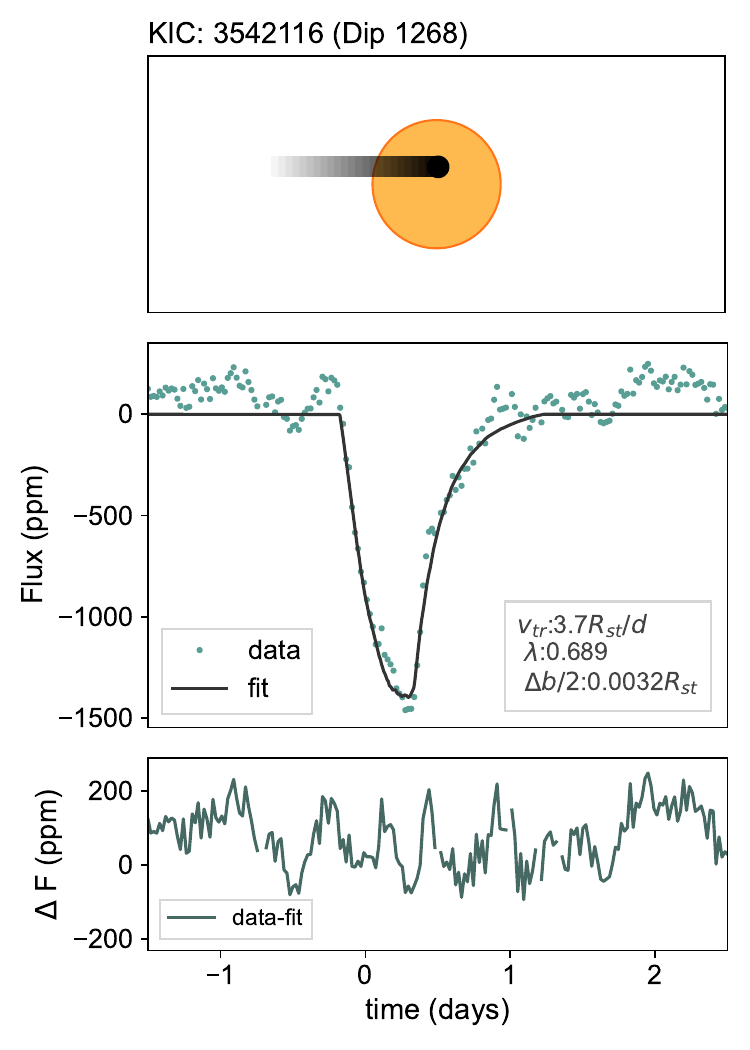}
\caption{Exocomet simulation for candidate KIC: 3542116 Dip 1268. The top panel shows a schematic of the simulation geometry. The comet size has been exaggerated for visual clarity. The middle panel shows the light curve of KIC 3542116 and the simulated comet model. The bottom panel shows the residuals (data - simulation). The fitted parameters are given.
\label{fig:comets}}
\end{figure}

For our simulation, we take a spherical coma and a dust tail of the comet to be of the same width. For the dust tail, we model an elongated rectangular shape with exponentially decreasing opacity. We take the normalization constant as $1$ and instead take the size of the comet system as a free parameter. Since any eccentricity information is unavailable, we assume this shape to be in a circular orbit around the star. The simulation agrees with the analytical model within the transit duration, but this assumption will not be valid out of transit. The schematic of our simulation geometry is shown in the top panel of Fig. \ref{fig:comets}. The simulation requires the free parameters: $R_{orb}, R_{comet}, \lambda, b, t_0$.
%____________________________________________________________________________
%___________________________________________________________________________

%_________________________________________________________________
\begin{figure*}[ht]
\plotone{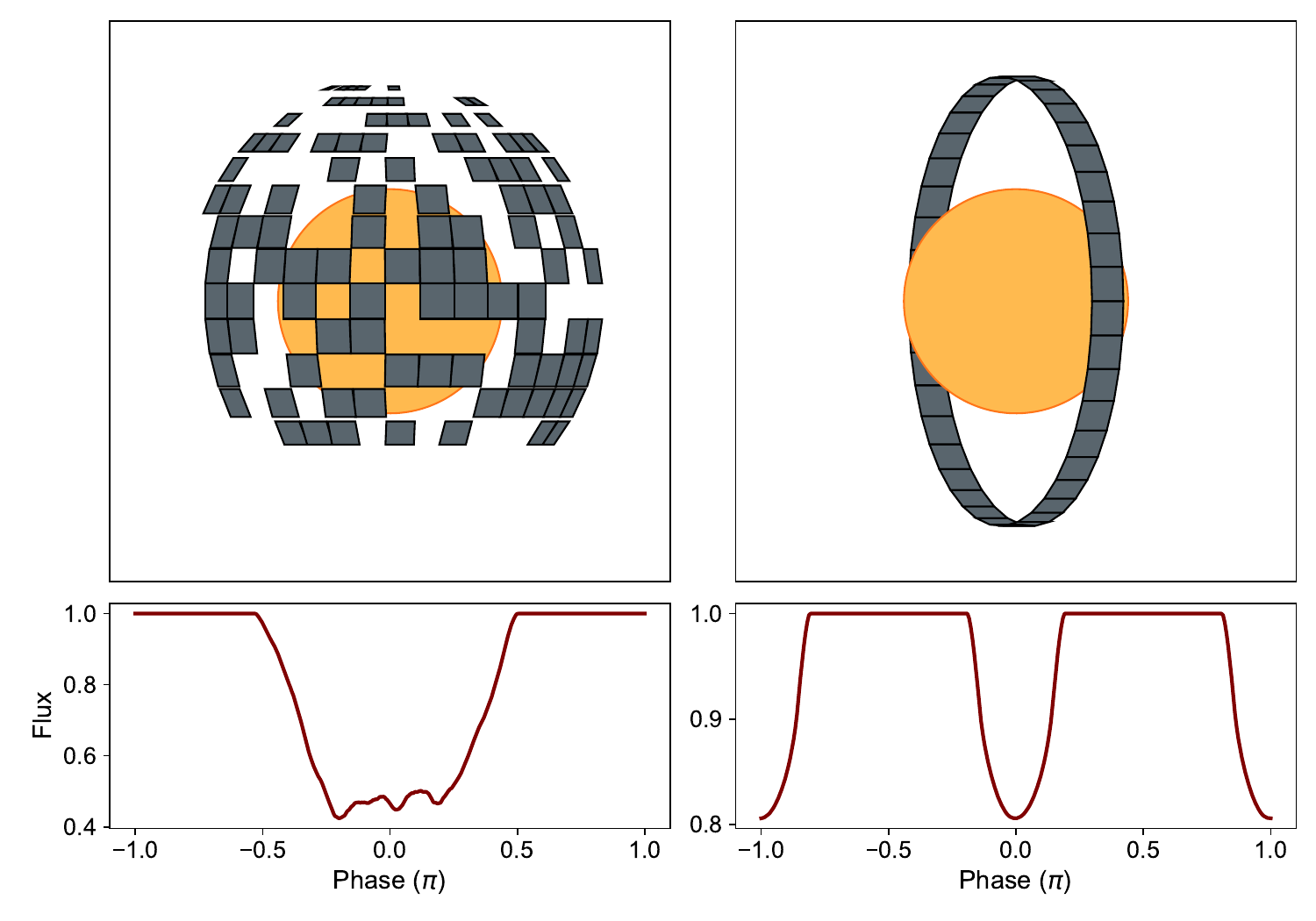}
\caption{Transit of weird objects. The left panel imagines a Dyson sphere in partial construction, causing gaps between the Dyson disks. The size of individual panels is $0.157 R_{st}$ at a distance of $2 R_{st}$ which corresponds to $40$ panels at the equatorial circumference. Away from the equator, the orbits and sizes are scaled according to the latitude, in order to maintain $40$ panels in any circumference. A total of $100$ panels are placed in this incomplete Dyson swarm. The bottom-left panel shows the resulting transit light curve, which shows a highly asymmetric and distorted geometry. The right panel shows a transit of a hypothetical polar ring. The ring is made of $41$ panels, each of size $0.153 R_{st}$ at a distance of $2 R_{st}$. The bottom-right panel shows the resulting transit, which is very similar to a planet transit with half the orbital period. 
\label{fig:aliens}}
\end{figure*}
%_________________________________________________________________

We simulate the light curve of one of the observed dips in the system KIC 3542116, which is observed at day $t_0 = 1268$. The impact parameter value is $0.27$ and is taken from the parameter obtained by \cite{exocomet_rappaport}. We use $u_1 = 0.2736$ and $u_2 = 0.3001$ as the quadratic limb darkening coefficients (see Eq.~\ref{eq:ldquad}) from the Vizier Catalog \citep{vizier} given for this star. 
Since our simulation gives the output flux in terms of the phase instead of time, we choose $ \Delta \theta = \Delta t (days) $, i.e. one orbital period as $2\pi$ days, which gives us,
$$v_t = R_{orb} (R_{st}/day).$$

The normalized light curve from Kepler data and the simulated light curve are shown in the middle panel of Fig. \ref{fig:comets}. We see that the simulation is in close agreement with the data. This is also visually similar to the fit demonstrated in Figure 3 of \cite{exocomet_rappaport}. We obtain a simulated value of $v_t = 3.7 R_{st}/day$ and $\lambda = 0.689$, which is close to the values mentioned in Table 2 of \cite{exocomet_rappaport}. We also obtain $R_{comet} = 0.0032 R_{st}$. The bottom panels of Fig. \ref{fig:comets} show the residuals of the fit, which is significantly lower than the noise in the transit light curves particularly away from the transit dip. This exercise highlights that \texttt{Yuti} can be used to model the exocomet light curves reliably. Moreover, the agreement between our simple visual good-fit and the parameters obtained by \cite{exocomet_rappaport} shows that our simulator can be reliably used to infer the physical parameters of the observed exocomet light curve.

%####################################################################################

\section{Transiting Megastructures: Dyson Swarm, Rings and Discs} \label{sec:AlienTransit}

%_________________________________________________________________
\begin{figure}[ht]
\plotone{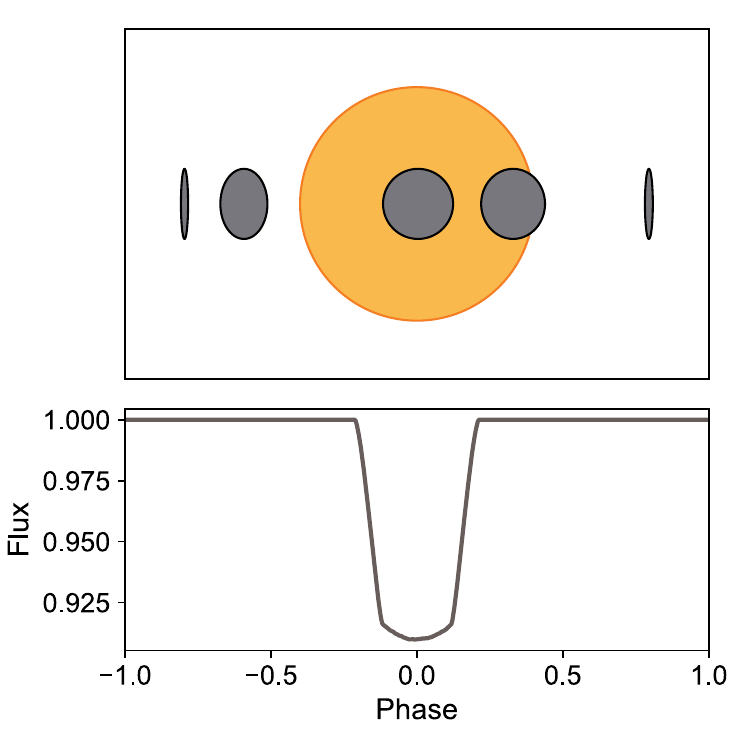}
\caption{A schematic showing the orbital configuration of a Dyson disk. The top panel shows five orientations of a flat circular 2D structure along the orbit. The size of the structure is $0.3 R_{st}$ at a distance of $2 R_{st}$. The bottom panel shows the resulting transit.
\label{fig:Dysonanim}}
\end{figure}

\begin{figure*}[ht]
\plotone{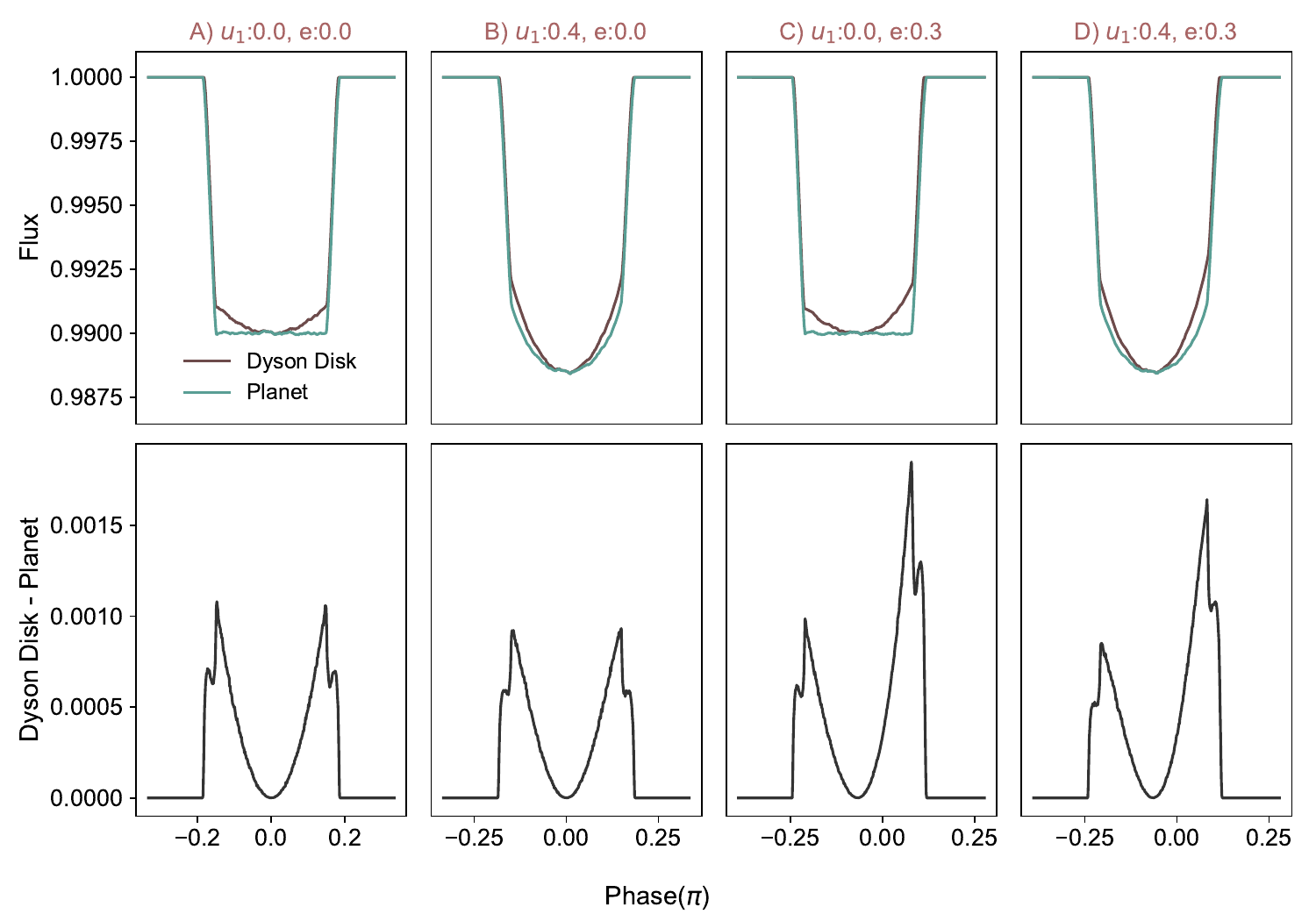}
\caption{Transit of a Dyson disk, compared to the transit of a planet of the same dimensions and distance to the star. The top panel shows the lightcurves obtained for a planet as well as Dyson disk transits. All the systems have a size $R_{pl}=0.1 R_{st}$ and $R_{orb}=2 R_{st}$. The bottom panel shows the residuals (Dyson Disk - Planet transit). Column A shows a simple transit with no eccentricity or limb darkening. While the planet transit is flat, the Dyson Disk transit shows an upward curvature during ingress and egress due to the change in the projection area. Column B adds limb darkening to the first transit which diminishes the residual feature.  Column C shows a transit with an eccentric orbit with the semi-major axis perpendicular to the line of sight. The Dyson Disk light curve shows significant asymmetry in ingress and egress due to the change of area of projection. Column D shows the same system with limb darkening. This results in a slightly diminished, but still significant residual feature.
\label{fig:2d3d}}
\end{figure*}
%________________________________________________________________
%

Building on this algorithm, we have extended the use of \texttt{Yuti} to investigate potential indicators of advanced alien civilizations through the transit of massive 
megastructures. Envisioned as creations of extraterrestrial societies to collect energy from their stars, these vast constructs are expected to produce distinctive 
patterns in the light curves. In the next subsection, we illustrate our simulator's capability to accurately model the light curves of a range of megastructures, 
with a specific focus on Dyson Swarm, Dyson Ring, and Dyson Disk - a circular opaque disk rotating around the host star - as prime examples.

\subsection{Dyson Swarm} \label{subsec:DysonSwarm}
The main example of a megastructure that can be built by a type 2 advanced civilization \citep{kardashev_org} is a Dyson sphere \citep{fDyson_pap}.
However, such a rigid structure as envisioned originally is not stable and instead one needs a collection of individual smaller structures, which is called a Dyson Swarm \citep{dysph_dyswm}.  For the Dyson swarm, one can imagine co-rotating solar panels in orbit around stars in such a way that they harness the energy of the host star. In order to simulate a Dyson Swarm in the process of building, we assemble large solar panels, one at a time, and allow them to revolve around a star. We can add more solar panels near the previous panels in such a way that eventually the whole area around the host star gets covered. One such incomplete structure is shown in the left-hand panel of Fig. \ref{fig:aliens}.  

We use this incomplete Dyson Swarm configuration in \texttt{Yuti} and obtain the light curve for it assuming it revolves around the star. In this Dyson Swarm, the orbit of the equatorial panels is at a distance $2R_{st}$ from the center of the star where $R_{st}$ denotes the radius of the star. To ensure seamless coverage around the circumference of the orbit, the size of each panel has been carefully calculated to allow for an exact integer number of panels to fit. Specifically in this example, the orbit's circumference is fully covered by 40 panels, each having a size of $0.157 \, R_{st}$ for the equatorial panels. For panels situated away from the equator, their orbital positioning is determined based on their latitude $\phi$, with orbits set at distances of $2cos\phi \, R_{st}$ from the star. This arrangement ensures that, regardless of latitude, the same number of panels (i.e. $40$) can be uniformly distributed around the circumference of each orbit. The simulation, depicted in Fig. \ref{fig:aliens}, incorporates a total of $100$ panels. The resulting light curve, presented in the bottom left panel of the figure, exhibits a significant transit dip due to the extensive coverage of the panels. Despite this coverage, there remain inter-panel gaps that allow light to pass through. As this incomplete Dyson sphere configuration orbits the star, the varying positions of these gaps and panels relative to the observer's line of sight produce the observable undulations seen in the transit light curve (see Fig. \ref{fig:aliens}).

In fact, the structure itself lacks stability in this configuration. It is assumed that panels rotate coherently along identical latitudinal lines which is an unstable Keplerian orbit except at $\phi = 0$.  However, this configuration was designed to demonstrate the versatility of our simulator in predicting light curves for any arbitrary configuration orbiting stars. In the next subsection, we turn our attention to a more stable construct: the Dyson ring. This structure comprises a singular ring of solar panels that rotates in unison around the host star.

\subsection{Dyson Ring} \label{subsec:DysonRing}

A complete Dyson sphere could also be envisioned as a series of rings. We imagine rings to be a made of numerous small connected panels. A complete equatorial ring covers the same area at all points in transit and would be undiscoverable by transit photometry. Therefore, we present a simulation of a hypothetical polar ring around the star. This is shown in the right panel of Figure \ref{fig:aliens}. The ring is assumed to be a rigid structure made of connected panels. The ring rotates around an axis which lies in the circular plane of the ring, and is aligned to the rotation axis of the star.

For the polar ring shown in Fig. \ref{fig:aliens} (right panel), we assembled $41$ square panels of the size $0.153 R_{\text{st}}$ orbiting at a distance of $2R_{\text{st}}$ from the star. The resulting light curve, displayed in the bottom right of Fig. \ref{fig:aliens}, is akin to the light curve from a planet transit, however, it occurs over half the orbital period.

\subsection{Dyson Disk} \label{subsec:SpaceMirror}
Transitioning from our investigation of Dyson swarms and rings, we explore a notable megastructure concept: a large solar panel in orbit, designed to constantly face the star for optimal energy collection. This raises the question: if this panel is circular and the size of a typical planet or larger, how does its transit light curve differ from that of a planet? We name such a panel as a Dyson disk. This can be envisioned as a part or component of a Dyson swarm. In addition to studying the light curve of the Dyson disks, in this subsection, we investigate the challenges associated with detecting those within the extensive dataset of the Kepler mission.

\subsubsection{Transit of a Dyson disk} \label{subsubsec:DyDscMain}
We expect the transiting Dyson disk will produce distinct signatures on the light curve, differing from those of a planet. It is because the projection of a planet on the sky plane is always a circle whereas the projection of the Dyson disk would be mostly elliptical except at the midpoint of the transit. This is depicted in the top panel of Fig.~\ref{fig:Dysonanim}. The requirement for Dyson disks to always face the star causes them to rotate, resulting in their predominantly elliptical projection on the sky plane. We illustrate the transit light curves for a Dyson disk, in the bottom panel of Fig.~\ref{fig:Dysonanim}, orbiting at the equator of a star (i.e., with an impact parameter of zero) at a distance of $2R_{\rm st}$ and with a size of $0.3R_{\rm st}$. 

For comparison, four panels in Fig.~\ref{fig:2d3d} show light curves generated for Dyson disks (at distance $2R_{\rm st}$, size of $0.1R_{\rm st}$ and $b=0$) for two different limb-darkening coefficients and two distinct eccentricities. Each panel also includes the light curve of a planet with identical radius and orbital parameters as the Dyson disk. In panel (A), we display the Dyson disk's transit around the star in a circular orbit (i.e. with $e=0$) without limb darkening. The lower segment of the transit reveals a curvature due to the gradual change in the area of the projected ellipse onto the star, contrasting with the planet's flat-bottomed light curve.

In the bottom part of the panel (A) of Fig.~\ref{fig:2d3d}, we subtract the Dyson disk's light curve from the planet's to plot a residual lightcurve. Notable features emerge in this comparison. Initially, the Dyson disk's entry is slightly delayed because of its rotation to maintain a star-facing orientation, resulting in a sharp increase in the residuals. As the Dyson disk enters the transit it starts to block the starlight, this modifies the slope of the residual light curve. A slight kink in the residual curve, preceding the slope change, marks the beginning of the Dyson disk's ingress. During the ingress phase, the planet obscures more light than the Dyson disk, causing the residuals to increase. Once the planet completes its ingress, the amount of light it blocks remains constant, while the Dyson disk continues to gradually increase its blockage. This reduces the residual light curves until the Dyson disk reaches the star's center, where its projection becomes a circle of the same size as the planet's projection, and both block an equal amount of radiation, nullifying the light curve difference at the transit's midpoint. Due to the symmetry of the situation, the pattern observed is mirrored until both the planet and Dyson disk exit the transit. 

The observed patterns in the light curves and their residual features persist for orbits around the star when realistic limb darkening is considered as shown in panel (B). However, the magnitude of the residuals is slightly diminished due to the limb-darkening effect, which dilutes the sharp transitions seen during ingress and egress because of otherwise uniformly bright edges.

Interestingly, in a specific setup within an eccentric orbit—where the periapsis is aligned perpendicular to the line of sight—the light curves for both the planet and Dyson disks display asymmetry, as demonstrated in panels (C) and (D) of Fig.~\ref{fig:2d3d}. This asymmetry primarily arises because both objects linger longer in the transit phase during egress. Furthermore, this discrepancy in the light curves is more pronounced on the asymmetric edge for the Dyson disk which is evident more clearly in the bottom panels where the difference pattern is also asymmetric, with one side showing a larger difference than the other.

The current model assumes Dyson Disks to be opaque 2D mirror like structures. However, a more detailed analysis will also include effects of reflected light, due to the shiny reflective surface of the mirror. This would be particularly evident during the eclipse of the Dyson Disk, and can lead to further interesting features, which can be explored in future versions of \texttt{Yuti}.

In order to detect these Dyson disks, one can estimate a crude signal-to-noise ratio (SNR) required in the transit data for determining the peaks in the residual features shown in the bottom panels of Fig.~\ref{fig:2d3d}. The SNR of a transit is defined as the ratio of the transit depth to the noise of the signal. For example, in order to get the optimal signal-to-noise ratio, we choose the transit shown in the second column of Fig.~\ref{fig:2d3d}, which has the shortest residual peak. For this, the transit depth is $1.158 \times 10^{-2}$, and the peak of the residual is at $9.323 \times 10^{-4}$. 

Therefore to detect this peak in the residual at a significance level of $3\sigma$, we need an SNR of $37$. This is consistent with many of the transits observed by the \emph{Kepler Space Telescope}, where $35\%$ of the detected Kepler Objects of interest have transits with SNR greater than $37$ \citep{archive}. However, the number estimated here is only applicable to a simulation of a Dyson disk with fixed parameters. In the following subsection, we explore what are the favourable conditions on the Dyson disk parameters that can increase the chances of their detection. 

\subsubsection{Viability of Detection} \label{subsubsec:DyDscViable}

%_______________________________________________________________________
\begin{figure*}[ht]
\plotone{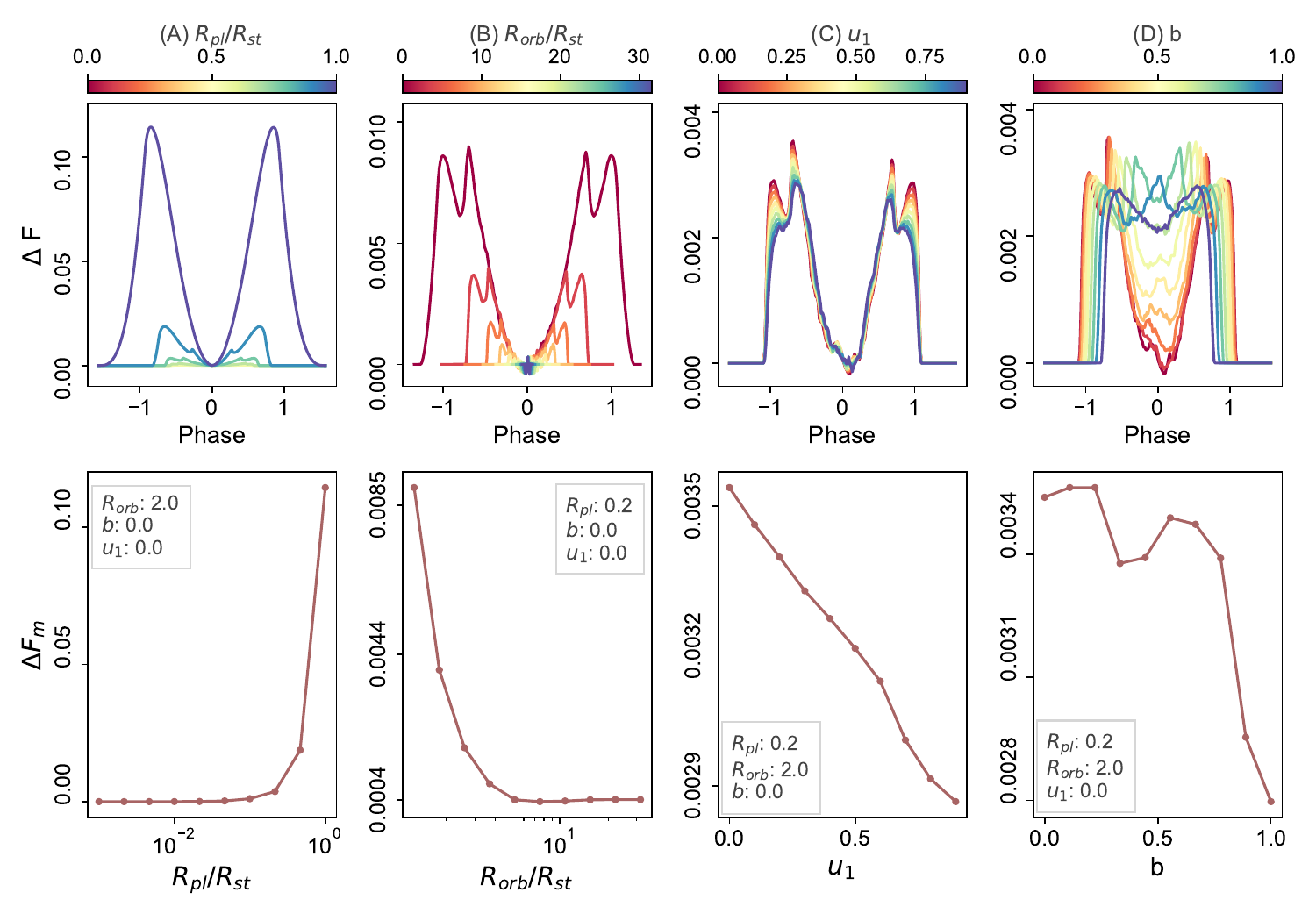}
\caption{Evolution of the residual feature obtained by Dyson Disk transit minus planet transit with transit parameters. Column (A) is for $R_{pl}/R_{st}$ . The top panel shows the residuals for different values of $R_{pl}/R_{st}$, which are represented by the color-map. The bottom panel shows the variation of $\Delta F_m$ with $R_{pl}/R_{st}$. $\Delta F_m$ is the maximum difference observed between the Dyson disk and planet light curve. There is a sharp increase in $\Delta F_m$ for values greater than $0.1$, which implies greater detectability. Column (B) is for $R_{orb}/R_{st}$. Once again, a sharp increase is observed for orbits closer than $10R_{st}$. Column (C) is for $u_1$. There is a linear decrease in $\Delta F_m$. Column (D) is for $b$. There is an overall decrease in $\Delta F_m$, however, the shapes of the residuals vary significantly, as is observed in the top panel.
\label{fig:2dtransitvar}}
\end{figure*}
%______________________________________________________________________

In order to examine the viability of the detection of Dyson disks, 
we first need to characterize the variation of the residual feature, i.e., the difference between the transit of a Dyson disk and a planet with the same size and orbital parameter (as shown in the bottom panels of Fig.~~\ref{fig:2d3d}).
We investigate these residuals for Dyson disk and planets with different transit parameters. For this, we generate simulated light curves of a Dyson disk and a planet for a range of transit parameters and calculate the residuals. Fig. \ref{fig:2dtransitvar} depicts the variation of the residual features across four 
parameters – the $R_{pl}/R_{st}$, ratio of the orbital radius $R_{orb}/R_{st}$, limb darkening parameter $u_1$ and the impact parameter $b$. The top panels show the different residual features. The color bars are indicative of the values of the transit parameters. For each simulated set, we find the maximum deviation in the residuals by calculating the magnitude of the peak of the residual feature ($\Delta F_m$). The bottom panels of Fig. \ref{fig:2dtransitvar} show the $\Delta F_m$ with the variation of the respective transit parameter.

Fig. \ref{fig:2dtransitvar} panel (A) shows the variation of residuals with $R_{pl}/R_{st}$. In figure the $R_{orb}/R_{st}$ is fixed at $2.0$ whereas $u_1, u_2$ and $b$ are kept at $0$. We can see that the $\Delta F_m$ value increases with the radius of the planet, especially in the regime of $R_{pl}>0.1 R_{st}$. 
This is expected, since the bigger the object, the more significant the distortion should be. Fig. \ref{fig:2dtransitvar} panel (B) shows the variation of residuals with $R_{orb}/R_{st}$. Here the $R_{pl}/R_{st}$ is kept at $0.2$ whereas $u_1, u_2$ and b are at $0$. We see that $\Delta F_m$ increases with decreasing radius of orbit. The distortion is significant for radii smaller than $10R_{st}$. This is because, due to a greater transit duration relative to the orbital period, the projection of a Dyson disk that is close to the star would have a large variation in the area of the projected elliptical shape during the transit duration, leading to greater differences in the residuals. 

Fig. \ref{fig:2dtransitvar} panel (C) shows the variation of residuals with limb darkening parameter $u_1$. Here $R_{orb}/R_{st}$ is kept at $2$, $R_{pl}/R_{st}$ at $0.2$ and $b$ as $0$. We observe that $\Delta F_m$ decreases linearly with an increase in the limb darkening parameter, which means that the chances of detection of such a feature may be better for stars with smaller limb darkening coefficients. 
As explained before, this is because the fainter edges of the stars dilute the contrast between the planet and the Dyson disk arising from the projection effect during ingress and egress.

Fig. \ref{fig:2dtransitvar} panel (D) shows the variation of residuals with impact parameter. Similar to previous case, here  $R_{orb}/R_{st}$ is kept at $2$, $R_{pl}/R_{st}$ at $0.2$ along with both $u_1$ and $u_2$ at $0$. We see an overall decline of $\Delta F_m$ with the impact parameter. However, this is more complex, because the shape of the residuals changes with impact parameter. At a higher inclination, the object never attains a full circular projection, causing a reduction in the transit depth, which is why the deviation at phase $0$ increases with $b$. 
However, the phase at which ingress occurs also reduces at high inclinations, leading to a lesser curved shape at ingress. This causes the sharp peak at the edges of the residuals to decline at high inclinations. 

From this analysis, we infer that the probability of observing deviations from typical transit signatures is greater for transits of Dyson disks that are large and revolving close to the star. To identify such features within existing datasets, targeting transits with $R_{pl}>0.1 R_{st}$ and $R_{orb} <10R_{st}$  would provide the highest chance of detection. Furthermore, focusing on transits characterized by low orbital inclinations and minimal limb darkening coefficients will 
improve detection capabilities.

%###########################################################################

\subsubsection{Degeneracies in Detections} \label{subsubsec:DyDscDeg}
%
%_________________________________________________________
\begin{figure*}[ht]
\plotone{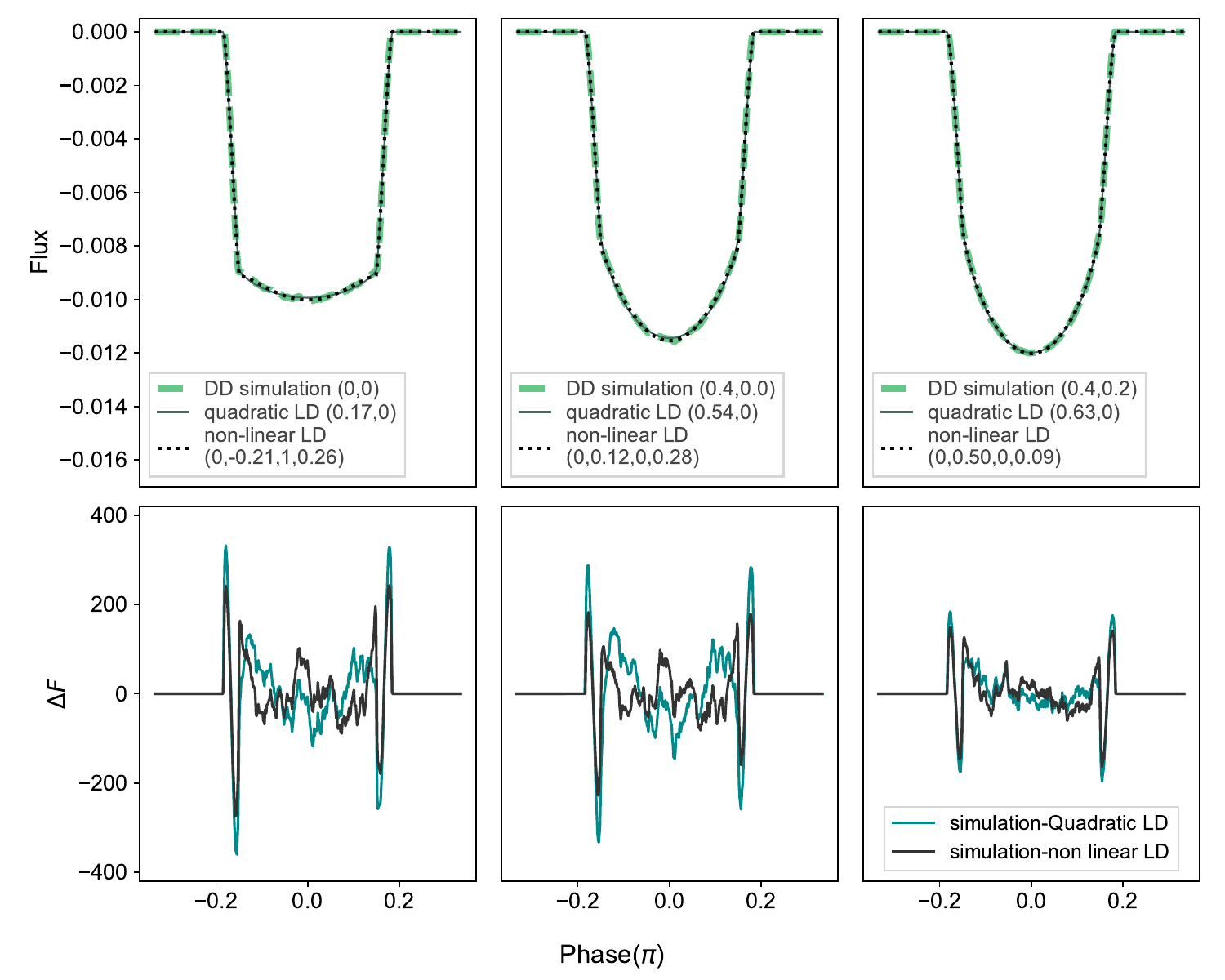}
\caption{Three simulated Dyson disk light curves with different limb-darkening parameters fitted with exoplanet models. The radius ($0.1 R_{st}$), orbit ($2 R_{st}$), and impact parameters ($b=0$) are kept the same for all three examples. A 3D planet model with both quadratic as well as non-linear limb darkening is fit to the Dyson Disk simulations. The top panels show the simulated light curves (green) and the fitted models. The bottom panels show the residual features (Simulation - Fit). The limb darkening coefficients for the simulated Dyson disk light curve as well as the planet fits are mentioned. From the residuals of the non-linear fit(purple) and that of the quadratic fit (blue), The non-linear law appears to fit slightly better than the quadratic law. Such a fit diminishes the residual feature, but kinks are observed at ingress and egress.
\label{fig:ldfit}}
\end{figure*}
%__________________________________________________________

The main challenge for finding megastructures such as the Dyson disk is to obtain an unambiguous detection. 
Specifically, for the Dyson disk, we investigate the potential for its transit signature to be indistinguishable from other natural phenomena.

Qualitatively, the transit of the Dyson disk appears quite similar to that of a planet, particularly due to the additional curvature at the bottom, which resembles a planet orbiting a star with pronounced limb darkening. Consequently, it seems reasonable to attempt fitting a transit light curve model of a planet to the transit of a Dyson disk. If limb darkening coefficients are considered variable parameters \citep{choice_ld}, in the absence of other independent measurements, 
fitting a planetary transit to the Dyson disk might result in higher limb darkening coefficients.  On the other hand, since the Dyson disk enters late in the transit reducing the transit duration will result in a fitted planet of a slightly 
smaller size and larger orbital radius than the Dyson disk. In this analysis, we explore the accuracy with which a planetary transit can be fit to the transit of the Dyson disk and whether there are distinct features in the fit capable of differentiating between the transit of the planet and a Dyson disk. 

In Fig. \ref{fig:ldfit}, we simulate Dyson disks of size $0.1 R_{\rm st}$ at an orbital distance of $2 R_{\rm st}$ with an impact parameter of zero 
for the stars with three limb darkening models. The first model  (\emph{left panel}) is without limb darkening and the second model (\emph{middle panel}) and the third model (\emph{right panel}) have different quadratic limb darkening coefficients namely $u_1 = 0.4, u_2 = 0.0$ for the middle panel and $u_1 = 0.4, u_2 = 0.2$ for the last panel. To these three models, we fit two models of planet light curves; one with quadratic limb darkening law (Eq. \ref{eq:ldquad}) while the other with a non-linear limb darkening law (Eq. \ref{eq:ldnlin}). We use a simple curve-fitting tool provided in Python library {\tt scipy} which uses the non-linear least squares method to fit parameters: $R_{pl}, R_{orb}, b, u_1$ and $u_2$.

As expected, the planetary transit fits give large limb darkening values, slightly smaller size ($R_{pl}$), and larger orbital radius $R_{orb}$. For example, as shown in the left panel of Fig.~\ref{fig:ldfit}, for the model with both limb darkening coefficients set to zero, the best-fit planet transit model with quadratic (non-linear) limb darkening law provides $u_1 =0.17$ while $u_2$ remains zero. This is accompanied by $R_{pl} = 0.0968$ and $R_{orb} = 2.009$. Whereas for the fit with the non-linear limb darkening model, our fits determine coefficients $(a,b,c,d) = (0.0, -0.21, 1.0, 0.26)$ along with $R_{pl} = 0.0971, R_{orb} = 2.009$.

A similar pattern of obtaining higher limb darkening coefficients is seen in the middle and right panels of Fig.~\ref{fig:ldfit}. For example, the middle panel contains a simulated light curve of the Dyson disk with $(u_1, u_2) = (0.4, 0)$ which is interpreted in the planet fit as $(u_1, u_2) = (0.54, 0)$ and $(R_{pl}, R_{orb}) = (0.0968, 1.997)$. The non linear limb darkening fit gives coefficients $(a,b,c,d) = (0.0, 0.12, 0.0, 0.28)$ and $(R_{pl}, R_{orb}) = (0.0974, 2.011)$. Whereas the third panel contains a simulated light curve of Dyson disk with $(u_1, u_2) = (0.4, 0.2)$ which is interpreted in the planet fit as $(u_1, u_2) = (0.63, 0), (R_{pl}, R_{orb}) = (0.0973, 2.006)$ using quadratic law and $(a,b,c,d) = (0.0, 0.50, 0.0, 0.86), (R_{pl}, R_{orb}) = (0.0975, 2.012)$ using non-linear limb darkening law.

Even though the Dyson disk appears to be degenerate with the limb darkening, it was ineffective in eliminating the residual features entirely.  It introduces kinks in the ingress and egress potentially because of trying to optimize for the size of the planet, orbital radius, and limb darkening to fit most of the transit. Using a higher-order or non-linear limb darkening law does improve the fit slightly but is not significant enough and still gives the same features in the residuals.

It is worth noting that when we fit the planet's transit to a model transit of a Dyson disk, not only do the residuals look different, but the value of peaks in the residuals also decreases. For example in the third panel of Fig. \ref{fig:2d3d}, the peak of residuals is at $9.32 \times 10^{-4}$ which is reduced by almost a factor of three (i.e $2.81 \times 10^{-4}$) when we fit the planet's transit in the middle panel of Fig. \ref{fig:ldfit}. Therefore this time we need a SNR of $123$, three times higher than what we estimated in Sec. \ref{subsubsec:DyDscMain}, for detecting such a signal in the residuals with a significance level of $3\sigma$.

Other degeneracies could arise from natural phenomena such as tidal distortion of planets, gravity-darkening, and star spots, which might be mistaken for Dyson disks.
We aim to thoroughly investigate these degeneracies in our future research as we search for Dyson disk candidates in Kepler and TESS datasets (Bhowmick et al. in prep.).

%################################################################################

\section{Conclusion}\label{sec:Summary}
We have developed a numerical transit simulator, \texttt{Yuti} designed to generate light curves for objects of any arbitrary shape transiting stars. \texttt{Yuti} employs a simple Monte-Carlo technique to compute accurate light curves, incorporating realistic limb-darkening profiles and geometries of stars. Throughout this article, we have demonstrated the broad applicability of the simulator to a variety of both natural and artificial structures in stellar orbits.

We demonstrate that \texttt{Yuti} accurately models the transits of single planets (Fig.~\ref{fig:plsim}) and multi-planetary systems such as Trappist-I (Fig. \ref{fig:trappist}). Additionally, it can model the transits of tidally distorted binary stars and giant planets (see Fig.~\ref{fig:tidewasp} and \ref{fig:hbtide}). 
Furthermore, we illustrate its capability to simulate transits of exocomets, using the example of a Kepler exocomet candidate (KIC 3452116, Fig.~\ref{fig:comets}). These examples collectively highlight the versatility of our simulator in generating light curves for a diverse range of natural transits. 

We show that \texttt{Yuti} is capable of generating transit light curves for artificial structures, such as those that might be constructed by Type I to Type II advanced civilizations, with examples including a Dyson swarm and a Dyson ring (Fig.~\ref{fig:aliens}). Additionally, we introduce a new structure that we name a Dyson disk: a large circular disk that rotates around a star while maintaining constant orientation towards it. Dyson disk can be thought of as a building block of a Dyson swarm. We explore how the transit light curve of a Dyson disk can be distinguished from that of a planet and discuss potential ambiguities due to limb darkening (Sec. \ref{subsec:SpaceMirror}). Furthermore, we provide a preliminary assessment of the signal-to-noise ratio required in the light curve to identify candidates for such Dyson disks. We will explore such Dyson disk candidates and potential degeneracy with other natural phenomena in the upcoming articles (Bhowmick et al. in prep.).

In conclusion, the versatility of our numerical transit simulator, \texttt{Yuti} will prove invaluable for understanding a wide array of transit phenomena involving both natural and artificial structures. The simulator will be made publicly available to facilitate the study of various natural transits and support the ongoing search for technosignatures, thereby aiding in the broader quest to detect signs of advanced civilizations.

\begin{acknowledgments}
This research has made use of the NASA Exoplanet Archive, which is operated by the California Institute of Technology, under contract with the National Aeronautics and Space Administration under the Exoplanet Exploration Program. This paper includes data collected by the Kepler mission and obtained from the MAST data archive at the Space Telescope Science Institute (STScI). STScI is operated by the Association of Universities for Research in Astronomy, Inc., under NASA contract NAS 5–26555. This paper also uses data from the program CH\_PR100013, which is a part of the public release of the WASP data as provided by the WASP consortium and services at the NASA Exoplanet Archive. For the various limb darkening coefficients, this work has made use of the VizieR catalog access tool, CDS, Strasbourg, France. 

Dr. Vikram Khaire acknowledges support for this work through the INSPIRE Faculty Award (No. DST/INSPIRE/04/2019/001580) of the Department of Science and Technology (DST), India. 
Ushasi Bhowmick thanks the Astronomy and Astrophysics Department, Indian Institute of Space Science and Technology for their extensive review of this research work, as well as Dr. A. S. Arya (SAC), Dr. Mehul R. Pandya (SAC),  Dr. Rashmi Sharma (SAC) and Director SAC for their support. 
We thank Nick Tusay for suggesting the megastructure name `Dyson disk' discussed in Section~\ref{subsec:SpaceMirror}. We also thank Shubhankar Gharote, Shivam Kumaran and Asif M. for suggesting the name `Yuti' for our transit simulator. Lastly, we would like to thank the anonymous referee for their helpful suggestions to improve the quality of the manuscript.

\end{acknowledgments}

\software{ellc \citep{ellc}}

\bibliography{paper}{}
\bibliographystyle{aasjournal}

\end{document}